\documentclass{aa} 
\usepackage{graphicx}
\usepackage{natbib}
\pdfoutput=1
\bibpunct{(}{)}{;}{a}{}{,}
\newcommand{\FIG}[1]{#1}

\begin{document}
\title{Faranoff-Riley type I jet deceleration at density discontinuities}
\subtitle{Relativistic hydrodynamics with realistic equation of state}

  \titlerunning{FR I Jet deceleration}
  \authorrunning{Z. Meliani et al.}
  \author{Z. Meliani \inst{3}, R. Keppens \inst{1,3,4}, B. Giacomazzo \inst{2}}

   \offprints{Z. Meliani}

\institute{FOM-Institute for Plasma Physics Rijnhuizen, Nieuwegein
\and Max-Planck-Institut f\"ur Gravitationsphysik, Albert-Einstein-Institut, Golm, Germany
\and Centre for Plasma Astrophysics, K.U.Leuven (Leuven Mathematical Modeling and Computational Science Center), Belgium
\and Astronomical Institute, Utrecht University, The Netherlands \\
 \email{Zakaria.Meliani@wis.kuleuven.be, Rony.Keppens@wis.kuleuven.be}
 }

   \date{Received ... / accepted ...}

\abstract{
The deceleration mechanisms for relativistic jets in active galactic nuclei 
remain an open question, and in this paper we propose a 
model which could explain sudden jet deceleration, invoking density discontinuities. This is particularly motivated by recent indications from
HYbrid MOrphology Radio Sources, suggesting that in some case Fanaroff-Riley 
classification is induced by variations in density of the external medium.
}
{Exploiting high resolution, numerical simulations, we demonstrate that for both high and low energy jets, always at high Lorentz factor,
a transition to a higher density environment can cause a significant fraction of the directed jet energy to be lost on reflection. This can explain
how one-sided jet deceleration and a transition to FR I type can occur in
HYbrid MOrphology Radio Sources, which start as FR II (and remain so on the other side).}
{For that purpose, we implemented in the relativistic 
hydrodynamic grid-adaptive AMRVAC code, the Synge-type equation of state
introduced in the general polytropic case by Meliani et al. (2004). To demonstrate its accuracy, we  set up various test 
problems in appendix, which we compare to exact solutions that we calculate as well.
We use the code to analyse the deceleration of jets in FR II/FR I radio galaxies, following 
them at high resolution across several hundreds of jet beam radii. 
}
{
We present results for 10 model computations, varying the inlet Lorentz factor from 10 to 20, including uniform or decreasing density profiles, and 
allowing for cylindrical versus conical jet models. As long as the jet propagates through uniform media, we find that the density contrast sets most of the propagation characteristics, fully consistent with previous modeling efforts. When the jet runs into a denser medium, we find a clear distinction in the decelaration of high energy jets 
depending on the encountered density jump. For fairly high density contrast, the jet becomes destabilised and compressed, decelerates strongly (up to subrelativistic speeds)
and can form knots. If the density contrast is too weak, the high energy jets continue with FR II characteristics.
The trend is similar for the low energy jet models, which start as underdense jets from the outset, and decelerate by entrainment in the lower region as well.
We point out differences that are found between cylindrical and conical jet models, together with dynamical details like the Richtmyer-Meshkov instabilities developing
at the original contact interface.
}
{}

\maketitle

\section{Introduction}
Accretion disks surrounding black holes, jets found in association with
compact objects, and Gamma Ray Bursts (GRBs) all represent violent astrophysical
phenomena. They are associated with relativistic flows, both with respect to the 
occuring velocities and to their prevailing equation of state.  
The jets from Active Galactic Nuclei (AGN) and in GRBs are  
accelerated in a short distance to reach a high Lorentz factor: typical values 
are $\gamma\sim 5-30$ \citep{Kellermannetal04} for AGNs and $\gamma>100$ for GRBs
\citep{Woods&Loeb95, Sari&Piran95}. In this acceleration 
phase, situated at the base of the jet, it is believed that jet energy is dominated by 
thermal energy and Poynting flux, and that a fraction of these energies 
contributes to the efficient acceleration of the flow. Thereby, the relativistic 
fluid changes its state from relativistic, corresponding to an effective 
polytropic index $\Gamma=4/3$, to classical (polytropic index $\Gamma=5/3$) when 
the thermal energy is converted to kinetic energy. Also further in the jet 
paths, where the jets interact with the surrounding medium, a fraction of the 
directed kinetic energy is converted to thermal energy at the shock fronts 
formed. According to the prevailing Lorentz factor of the jet, these shocks 
could be relativistic and therefore very efficient to convert kinetic to thermal 
energy, or could be Newtonian and only weakly efficient in compression. 
To investigate the occuring relativistic flows, a realistic equation of state 
must therefore be adopted, to handle both classical as well as relativistic temperature 
variations in space and time. For that purpose, we implemented in the relativistic 
(magneto)hydrodynamic grid-adaptive AMRVAC code~\citep{Melianietal07, Keppens03, vanderHolst07}, the Synge-type equation of state
introduced for a general polytropic case as in Meliani et al. (2004) and as applied in the adiabatic case by Mignone \& McKinney (2007). 
To demonstrate its accuracy, we set up various stringent test 
problems in an appendix, which we compare to exact solutions for Riemann problems, that we calculate as well.
The latter include Riemann problems at Lorentz factors of order $\gamma\approx 100$,
which represent extreme values relevant for Gamma Ray Burst flows.

The development of relativistic numerical hydrodynamic codes can help us 
understand the physics of astrophysical jet propagation. In the last decade, 
significant progress was made in numerical special relativistic hydrodynamic (HD)
and magnetohydrodynamic codes. Various authors worked on the 
development of conservative shock-capturing schemes which use either exact or 
approximate Riemann solver based methods for relativistic hydrodynamics 
\citep{Eulderink&Melemma94, Fontetal94, Aloyetal99, DelZanna&Bucciantini02, Mignone&Bodo05} 
(for a review see \cite{Marti&Muller03}).
The study of relativistic hydrodynamic fluids benefits also from using  
spatial and temporal adaptive techniques, or Adaptive Mesh Refinement \citep{Duncan&Hughes94, Zhang&MacFadyen06, Melianietal07, Wangetal07}.
The various numerical simulations usually involve a simplified equation of state 
(EOS) with a constant polytropic index. Notable exceptions with more realistic 
EOS treatments are found in 
\cite{Mignone05, Mignone&McKinney07, Perucho&Marti07}.
We present in the appendix to this paper the required extension of the AMRVAC code \citep{Melianietal07, Keppens03, vanderHolst07} with the more realistic polytropic EOS introduced by~\cite{Melianietal04} which is based on the Synge gas equation, the 
relativistically correct perfect gas law~\citep{Synge57, Mathews71}. 
The equations which we solve, and the schemes used, are also mentioned there.

Many previous investigations of relativistic jet propagation through the interstellar 
medium (ISM) in radio galaxies concentrate on uniform ISM conditions
\citep{Duncan&Hughes94, Martietal97, KomissarovetFall98, Aloyetal99}.
These investigations contributed to our understanding of the jet deceleration, 
where one then distinguishes dynamics
for Faranoff-Riley type I (FR I) and FR II radio galaxies, according
to the power of the jet and hence the accretion rate in their galactic center. 
In the FR I radio galaxies, the associated jets are 
relativistic on parsec scale and sub-relativistic on kiloparsec-scales, so that 
jet deceleration, and thus energy redistributions, must happen on kiloparsec 
scales \citep{Hardcastleet05}. Various studies have looked into possible deceleration processes with both non-relativistic and relativistic HD codes. \cite{Hooda&wiita96} and \cite{Hooda&wiita98} investigated with a classical HD
code the propagation of a 3D jet through an interface between 
dense ISM and less dense intercluster medium. They found that for such
interface marking a density decrease, the jet does not decelerate when crossing it, and that the
deceleration mostly happens in the lower region (dense ISM). Moreover, in their 
study the jets are not deflected at the interface ISM/ICM. \cite{Normanetal88} also
use a classical code to study the sudden deceleration of jets, when
the jet crosses a shock wave in the external medium.
Other works investigate the interaction of jets with dense clouds in both
non-relativistic hydrodynamic simulations \citep{Saxtonetal05} and 
relativistic simulations \citep{Choietal07}. 
\citet{Duncan&Hughes96} study relativistic jet propagation in uniform overdense media
and use an equation of state for a pair-plasma. 
\citet{Schecketal02} study the influence of the matter composition of a
high jet energy jet on its interaction with the external medium. They investigate the two extreme cases of pure leptonic and baryonic plasmas.
\citet{Perucho&Marti07} examined the propagation of the low 
energy jet of the specific FR I radio source 3C 31 using an elaborate hydrodynamics model, with
an equation of state that distinguishes the contribution of leptons and baryons to density and pressure.
Most recently, \citet{Rossietal08} investigate 
the propagation of jets in uniform overdense media in 3D, using a realistic synge EOS. 

\begin{table*}
\caption{The most relevant characteristics and parameters for various selected relativistic hydrodynamic simulations in the literature:
DH94 \citep{Duncan&Hughes94}, M97 \citep{Martietal97}, H98 \citep{Hardeeetal98}, R99~\citep{Rosenetal99}
S02~\citep{Schecketal02},  PM07 \citep{Perucho&Marti07}, R08 \citep{Rossietal08}. Our simulations are indicated by MKG. We mention the type of the simulation 2D or 3D and use of AMR, jet beam kinetic luminosity $L_{\rm b}$ (when available), $\gamma_{\rm b}$ Lorentz factor of the beam, $M_{\rm b}$ relativistic Mach number, $\eta$ density ratio (``d" means density variation), $\theta$ open angle of the jet, $R_{\rm b}/r_{\rm cell}$ grid cells through jet beam, type of EOS ("Lep" means Leptonic, "Bar" Baryonic), endtime of the simulation in units of light crossing time of the jet beam radius.}
\label{table:1}      
\centering                                      
\begin{tabular}{c c c c c c c c c c}       
\hline\hline                        
Paper & case & $L_{\rm b}$ & $\gamma_{\rm b}$ & $M_{\rm b}$ & $\rho_{\rm medium}/\rho_{\rm jet}$& $\theta_{\rm b} $& $R_{\rm b}/r_{\rm cell} $ &EOS& Size in $R_{\rm b}$ (and time)\\    
\hline                                   
\hline                                   
DH94/R99  & A (2D, AMR)  & - &1.048 & 6 & $10.0$&0& 24& $5/3$&  $10 \times 41.67 $ (-)\\
DH94/R99  & B (2D, AMR)  & - &5.0 & 8 & $10.0$&0& 24& $5/3$&  $16.67 \times  41.67$ (-)\\
DH94/R99  & C (2D, AMR)  & - &10.0 & 8 & $10.0$ &0& 24& $5/3$& $16.67\times 41.67 $ (-)\\
DH94/R99  & D (2D, AMR)  & - &10.0 & 15 & $10.0$ &0& 24& $4/3$&  $16.67 \times 41.67$ (-)\\
M97  & A1 (2D)  & - &7.1 & 9.97 & $10^2$ &0& 20& $4/3$&  $10.5 \times 50$ (50)\\
M97  & A2 (2D)  & - &22.37 & 31.86 & $10^2$ &0& 20& $4/3$&  $10.5 \times 50$ (50)\\
M97  & a1 (2D)  & - &7.1 & 9.97 & $1.0$ &0& 20& $4/3$& $10.5 \times 25$ (25)\\
M97  & a2 (2D)  & - &7.1 & 9.97 & $10.0$ &0& 20& $4/3$& $10.5 \times 25$ (25)\\
M97  & B1 (2D)  & - &7.1 & 41.95 & $10^2$ &0& 20& $4/3$&   $10.5 \times 50$ (50)\\
M97  & B2 (2D)  & - &22.37 & 132.32 & $10^2$ &0& 20& $4/3$&   $10.5 \times 50$ (50)\\
M97  & b1 (2D)  & - &2.29 & 13.61 & $10.0$ &0& 20& $4/3$&  $10.5 \times 25$ (25)\\
M97  & b2 (2D)  & - &7.1 & 41.95 & $10.0$ &0& 20& $4/3$& $10.5 \times 25$ (25)\\
M97  & C1 (2D)  & - &2.29 & 13.61 & $10^2$ &0& 20& $5/3$&  $10.5 \times 50$ (50)\\
M97  & C2 (2D)  & - &7.1 & 41.95 & $10^2$ &0& 20& $5/3$& $10.5 \times 50$ (50)\\
M97  & C3 (2D)  & - &22.37 & 132.32 & $10^2$ &0& 20& $5/3$&   $10.5 \times 50$ (50)\\
M97  & c1 (2D)  & - &2.29 & 13.61 & $10.0$ &0& 20& $5/3$&  $10.5 \times 25$ (25)\\
M97  & c2 (2D)  & - &7.1 & 41.95 & $10.0$ &0& 20& $5/3$&  $10.5 \times 25$ (25)\\
H98 & B (2D, AMR)   & - &5.52 & 8.87 & $10.0$ &0& 24& $5/3$& $16.67 \times  41.67$ (140.0)\\
H98 & C (2D, AMR)   & - &14.35 & 11.52 & $10.0$ &0& 24& $5/3$& $16.67 \times  41.67$ (21.43)\\
H98 & D (2D, AMR)   & - &10.0 & 15.0 & $10.0$ &0& 24& $4/3$&  $16.67 \times  41.67$ (35.273)\\
H98 & E (2D, AMR)   & - &2.55 & 8.53 & $10.0$ &0& 24& $5/3$& $16.67 \times  41.67$ (140.0)\\
S02 & A (2D)  & $10^{46}$ &6.62 & 9.24 & $10^5$ &0& 6& Lep& $200\times 500$ (5950)\\
S02 & B (2D)  & $10^{46}$ &6.62 & 14.33 & $10^3$ &0& 6& Lep& $200\times 500$ (5400)\\
S02 & C (2D)  & $10^{46}$ &7.95 & 130.14 & $10^3$ &0& 6& Bar& $200\times 500$ (5300)\\
PM07 & A (2D)  & $10^{44}$ &2.0 & 4.75 & $10^5$ ``d" &0& 16& Lep/Bar& $200\times 450$ (37231)\\
R08& A (3D)  & - &10.0 & 28.3 & $10^2$ &0& 20& Synge&  $50 \times 150 \times 50$ (240)\\
R08& B (3D)   & - &10.0 & 28.3 & $10^4$ &0& 20& Synge& $60 \times 75 \times 60$  (760)\\
R08& C (3D)   & - &10.0 & 28.3 & $10^4$ &0& 12& Synge&  $50 \times 75\times 50$ (760)\\
R08& D (3D)   & - &10.0 & 300 & $10^4$ &0& 20& Synge&  $50 \times 150\times 50$ (760)\\
R08& E (3D)   & - &10.0 & 300 & $10^2$ &0& 12& Synge&  $24 \times 200\times 24$ (150)\\
MKG & A (2D, AMR)   & $10^{46}$ &20.0 & 1200 & $0.1496$-$4.687$  &0& 64& Synge& $10 \times 400$ (380)\\
MKG & B (2D, AMR)  &  $10^{46}$ &20.0 & 1200 & $0.1496$-$671.22 $ &0& 128& Synge& $10 \times 400$ (900)\\
MKG & C (2D, AMR)   & $10^{43}$ &10.0 & 39 & $36.52$-$203.66$  ``d" &0& 76& Synge& $10 \times 400$ (820)\\
MKG & D (2D, AMR)   & $10^{43}$ &10.0 & 39 & $36.52$-$148.15$  ``d"  &1& 76& Synge& $10 \times 400$ (820)\\
MKG & E (2D, AMR)   & $10^{43}$ &20.0 & 35 & $146$-$847.46$ ``d" &0& 76& Synge& $40 \times 400$ (820)\\
MKG & F (2D, AMR)   & $10^{43}$ &20.0 & 35 & $146$-$645.16$   ``d" &1& 76& Synge& $40 \times 400$ (820)\\
MKG & G (2D, AMR)   & $10^{46}$ &10.0 & 1300 & $0.033$-$0.495$   ``d" &0& 144& Synge& $40 \times 400$ (380)\\
MKG & H (2D, AMR)   & $10^{46}$ &10.0 & 1300 & $0.033$-$30.6748$  ``d"  &1& 76& Synge& $40 \times 400$ (800)\\
MKG & I (2D, AMR)   & $10^{46}$ &10.0 & 1300 & $0.033$-$0.306748$  ``d"  &1& 76& Synge& $40 \times 400$ (480)\\
MKG & J (2D, AMR)   & $10^{46}$ &20.0 & 1200 & $0.1496$-$12.95$  ``d" &1& 76& Synge& $40 \times 400$ (480)\\
\hline                                             
\end{tabular}
\end{table*}
In this paper, we investigate a scenario of relativistic jet 
propagation through an ISM with a sudden jump in density. 
Since jets propagate over enormous distances, it is inevitable that they encounter 
density jumps in the external medium. 
As we assume that matter in the inner part of the radio galaxy has been cleared away during the evolution of the radio galaxy,
we typically follow jet dynamics through a lighter medium at first, which then suddenly changes far away to a denser medium. In a sequence of model runs, we vary the jet and medium properties to cover both high and low jet beam kinetic luminosities, straight as well as conical jets (varying the jet opening angle), and allow for either uniform or decreasing density profiles. To put our work into perspective and
to appreciate its relation to previous works better, we compiled in Table~\ref{table:1} the most relevant parameters and simulation
specifics for selected relativistic hydro models. The most novel assumption is the transition from low to high density across
a contact interface (across which the pressure is necessarily constant), which is different from the work done by \citet{Hooda&wiita98} 
where they use a classical code with a density decrease across the contact, and also different from the study by
\citet{Lokenetal93}, where they use a classical code to look into jet propagation through a pressure wall.   
We explore the influence of a sudden density jump in ISM on jet propagation, stability, 
and formation of the bow shock. We aim to better understand the 
jet efficiency in transporting energy and mass from the central AGN regions to the denser ISM at large distances, where the jet cocoon gets formed. 
Note also from Table~\ref{table:1} that our models extend the previous works most notably by typically covering much larger distances than previously studied (as we
need to obtain representative endstate morphologies in both lower and upper media), and are invariably in the high Lorentz factor regime, combined with a high resolution through the jet beam.
The high energy jets we consider start out as denser than their immediate surroundings (a likely property deduced from all common jet launch scenarios, but previously ignored by all jet simulations to date). Moreover, the density contrasts we investigate between
jet and outer medium are much more reasonable than the 5 orders of magnitude density differences previously used by~\citet{Schecketal02}.
In what follows, we first motivate our model assumptions and the simulation setup in Section~\ref{secmodel}. We discuss our main findings
in Section~\ref{secres}.

\section{Relativistic jet propagation and deceleration}\label{secmodel}

\subsection{Motivation}
The radio loud Fanaroff-Riley galaxies \citep{Fanaroff&Riley74} have extensively been studied in the last decade, because their jets show intriguing behaviour. 
In fact, radio loud galaxies are grouped in two main classes according to their radio map morphology. A typical FR I is brighter near the center and fades out towards the edge, whereas FR II are brightest at the edges and fainter toward the center. 
\cite{Fanaroff&Riley74} discovered that the break in FR I/FR II occurs around the radio luminosity of 
$P_{178 {\rm MHz}}\sim10^{25} {\rm W Hz^{-1} sr^{-1}}$, with almost all sources below the break value being of type FR I.
In contrast to FR II galaxies where the jet remains relativistic and narrow in 
all scales, the jet in FR I \citep{Giovanniniet05} is relativistic at the parsec-scale \citep{Bridle92} 
and in many cases subrelativistic and diffuse in kparsec-scales \citep{Giovanniniet01}. 
The FR I jet shows side to side asymmetries which decrease
with distance from the central engine, caused by the decrease of Doppler 
beaming in an intrinsically symmetrical, decelerating jet \citep{Laingetal99}.
However, in some FR I the structure of the jet in the kiloparsec scale appears more complicated, with an inner spine which remains 
relativistic and an outer shell that decelerates and becomes sub-relativistic \citep{Canvinetal05}. 

It is commonly accepted that the morphological differences between FR I/FR II 
arise because of differences in the physical conditions of 
jet interaction with its environment. However, it is still under debate whether the observed deceleration in FR I 
jets is related to the mechanism of jet launching and thus to the properties of the central engine in AGN
\citep{Ghisellini&Celotti01, Kaiser&Best07}, or 
rather related to the properties of the external medium: the host galaxy 
\citep{Zirbel97b} and the circumgalactic gas 
\citep{DeYoung93, Kaiser&Alexander97}. 
Some authors \citet{Woldetal07} claim that the FR I/FR II dichotomy 
is caused by the combination of a central engine factor, with a contribution of the external medium.
Observational evidence for the external medium influence comes from
the existence of a group of peculiar ``HYbrid MOrphology Radio Source" (HYMORS), 
pointed out by \citet{GobalKtishna&Witta02}.
These radio sources appear to have an FR II type on one side and an FR I type diffuse radio lobe on 
the other side of the active nucleus. 
This supports the idea that the different Fanaroff-Riley morphologies are attributed in some cases to the properties of the ambient media, since for HYMORS, similar jets (power, composition, Lorentz factor) likely emerge from the central engine on each side.
\cite{Heywood&Blundell&Rawlings07} analysed radio images for a set of quasars 7C 
(radiosource galaxies) and found that some FR I seem to have a radio luminosity 
exceeding the original FR I/FR II dividing line. They also confirmed the existence of HYMORS and suggested that these sources have high power jets (typical for FR II) and they yield FR I or FR II morphologies according to their external medium. In a rarefied medium, the jet is 
`laminar' and remains collimated all the way to the intergalactic medium (IGM) where the hotspot forms (the impact site of the jet in the 
IGM). This gives an FR II jet morphology with the typical ``lobe". In dense or clumpy medium, the jet interacts strongly with its 
external medium and dissipates its energy more gradually as it propagates in this dense medium. This gives the FR I jet morphology with the typical ``plume" at large spatial scales. 

We point out several concrete examples next.
The bright one-sided (FR I-like), diffuse jet in the radio galaxy 3C 321 (which is an FR II radio source) is an excellent example of a HYMORS.  
\citet{Evansetal08} argue that the one-sided diffuse jet is the result of an interaction of the jet with the companion galaxy. 
The fact that in these objects, both sides of the jets are relativistic and stable at the parsec scale, and the difference appears at the large scale, means that the variation in the external medium must occur at some distance from the central engine. This is what we assume in our models.
The radio galaxy Cen A shows also a difference in the radio morphology between the two sides of the jet as it propagates on kiloparsec scales, with edge-brightened lobes (FR I-like) on one side, and on the other side a central (fine structured) lobe (FR II-like) \citep{Kraftetal03}. Yet another example is the powerful radio source Hercules A (3C 348) which has a jet kinetic luminosity $10^{46} {\rm ergs/s}$ \citep{McNamaraetal05} and exhibits a mixed FR I/FR II morphology \citep{SadunetMorrison02}. These observations confirm the idea that in Cen A and Hercules A (3C 348), the external medium plays a key role in the appearance of the jet and its dynamics.
There is also evidence for disruptions or variations of radio morphologies induced by inhomogeneous 
medium. For example the evident hierarchical structure of M87 in its radio image \citep{Owenetal00},
with the possibility of the existence of two halos: an inner halo that could be with more porous 
structure, while the jet interacts mainly with the outer halo \citep{Owenetal00}.
In conclusion, some FR I jets propagate through clumpy \citep{Croftetal06} or dense media, evidently encountering sudden density changes. This is known to give rise to a strong interaction, and the jet loses its energy by 
entrainment and diffusion \citep{DeYoung93, DeYoung96, Rosen&Hardee00, Tavecchioetal06} and
 also forms knots along the jet \citep{Owen89}. 
We now detail our new, more elaborate model, with which we aim to explain observations of jets in 
radio galaxies that are relativistic on small scale and decelerate to 
sub-relativistic velocities going to the large scale.

\subsection{Model}
To make the problem more tractable, we assume axisymmetry and we neglect the 
influence of the magnetic field in the dynamics. In our scenario, we assume 
the existence of a density jump in the host 
galaxy or in circumgalactic gas which could be responsible for the
jet deceleration and knot formation in the asymptotic region. Since the jets travel for enormous distances, it is
inevitable that they encounter various sudden transitions of interstellar medium
properties. These could be traveling shock fronts, more gradually varying background
variations as one traverses regions of differing gravitational potential, or
contact discontinuities indicative of boundaries between varying regions of influence.
We concentrate on the latter, representing density (and entropy) changes across which
total force balance holds (uniform pressure), as these are invariably found in any kind
of hydrodynamic interaction involving winds, outflows, etc.

We model jet propagation through two distinct
media. In the first part, a low and uniform density is assumed, such that in 
this region the jet-external medium interaction is weak. We will typically consider jets that are denser than the lower surrounding medium
(but also include models where the jet is already underdense in this region). 
If the external medium in the inner region is denser than the jet, 
one expects strong disruption of the jet and hence
diffuse and destabilised jets in this inner (parsec scale) region where the jet would 
decelerate and dissipate its energy. This is in contrast with 
observations of narrow and relativistic jets in the inner region of FR I 
galaxies and in the HYMORS, where the energy deposited at the large-scale is comparable to the energy of the central engine~\citep{Rawlings&Saunders91}.
In this work, we are particularly interested in radio loud FR I 
with a powerful jet at high Lorentz factor in the lower scale. This is the case
in the HYMORS, where the jets have the same properties at small scale than the 
high energy FR II jets. The differences appear only in one side of the jet at the larger scale, 
where the jet
morphology changes to show structure characteristic of an FR I. 
Therefore, we assume that further downstream, the jet encounters a high density medium, such that 
the jet undergoes a strong 
deceleration and a strong compression. 
This latter part is typically the only part simulated in previous numerical 
jet studies, where a prescribed
jet configuration penetrates a usually uniform, high density, external medium. 
Moreover, we study the effect of the initial opening angle of the jet in its 
interaction with (layered or stratified) external medium.

For the density jump in the case of 3C 321, we can think of the lower region as the rarefied medium in the inter-cluster medium, while the upper region represents denser medium of the companion galaxy. In other HYMORS, the density jump is then thought to occur on one side of the AGN where there are denser molecular clouds in the inter-cluster environment or interstellar medium. 
One of the main points we hereby address for the first time is the change in jet 
head properties during the propagation phase in the inner region, and how this in turn
affects the jet stability in the upper, denser medium. Indeed, the
jet interacts with the inner medium mainly through the bow shock, 
whereas the jet beam is only weakly disturbed laterally. 
The shocked swept-up matter during this phase, and the shocked part of 
the beam, all constitute a structured bow shock ahead of the beam, and both 
shocked regions form a new hot layer with lower Lorentz factor, characterised 
by lower Mach number. The interaction of 
this preformed, structured jet head with denser external medium in the outer region should 
produce a strong cocoon and backflow. This latter will also disturb the 
non-shocked jet beam as it penetrates the denser medium. This will increase the entrainment of 
ambient material through velocity shear instabilities and decelerate the jet 
\citep{DeYoung93, Perucho&Marti07}.

\subsection{Initial conditions}

Deducing the precise internal properties for an FR I jet and its environment from 
observations is a difficult task, and density contrasts in particular have partly been obtained on the basis of numerical studies. However, from generic 
properties of jet propagation in various FR I galaxies, 
one can estimate the kinetic luminosity and the jet beam 
Lorentz factor. Then, the choice of jet and environment parameters is determined on 
input by the kinetic luminosity of the jet and the estimated jet propagation 
speed in the two media. The kinetic luminosity of a typical powerful jet is 
$L_{\rm jet,Kin} \sim 10^{46} \,{\rm ergs/s}$~ \citep{Rawlings&Saunders91, Daly95, Carilli&Barthel96, Wanetal00, Drake03, Tavescchioetal04, Kino&Takahara04} and we used this observationally supported value for the simulations indicated as cases A, B, G, H, I, J (see Table~\ref{table:1}). 
We investigate also the propagation of low energy jets and compare them with these more powerful jets.
For more radio quiet galaxies, a lower energy jet with $L_{\rm jet,Kin} \sim 10^{43} \,{\rm ergs/s}$ \citep{Allenetal06} is deduced and this value is used 
for the simulations C, D, E, F. The integrated energy flux over the beam cross section is computed from   
(e.g. \cite{Bicknell&Begelman96, Martietal97, Rosenetal99, Schecketal02})
\begin{equation}\label{KLflux}
L_{\rm jet,Kin}= \left(\gamma_{\rm b}\,h_{\rm b}-1\right)\rho_{\rm b}\gamma_{\rm b}\pi
\,R_{\rm b}^2 v_{\rm b}\,,
\end{equation}
where the subscript ``$\rm b$" indicates the jet beam, $\rho_{\rm b}$ is 
its density, $\gamma_{\rm b}$ is the Lorentz factor, $v_{\rm b}$ is the speed,
$\rho_{\rm b} h_{\rm b} = \rho_{\rm b} + \frac{\Gamma}{\Gamma-1}\,p_{\rm b}$ is the 
enthalpy, and $R_{\rm b}$ is the jet radius. For the latter, when we assume a jet with opening angle of $\theta_{\rm b}=3^{o}$ at $1 \, {\rm pc}$, the observed value in Centaurus A \citep{Horiuchietal06}, we get a jet radius $R_{\rm b} \sim 0.05 {\rm pc}$.
We will fix the radius $R_{\rm b}=0.05 {\rm pc}$ on our jet input boundary, which in turn is assumed to be located at a distance from the source of $0.5{\rm pc}$.

\begin{figure*}
\begin{center}
\FIG{
{\rotatebox{90}{\resizebox{5cm}{\textwidth}{\includegraphics{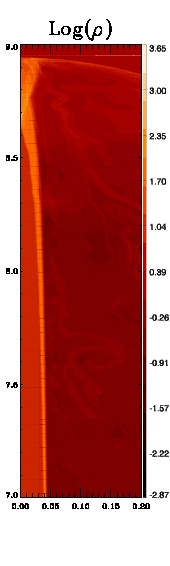}}}}
{\rotatebox{90}{\resizebox{5cm}{\textwidth}{\includegraphics{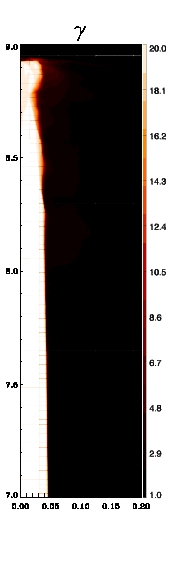}}}}
{\rotatebox{90}{\resizebox{5cm}{\textwidth}{\includegraphics{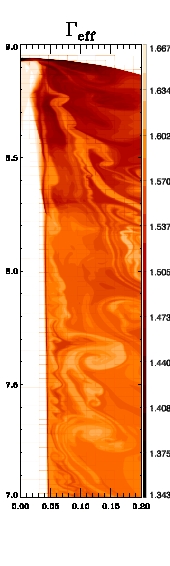}}}}
}
\caption{A zoom on the jet head at $t=160$ light crossing times of the jet radius $R_{\rm b}$, for cases A, B and J,
just prior to penetrating the denser upper medium. Shown are: (Top) the logarithm of density, (middle) the Lorentz factor, (bottom) 
the effective polytropic index.
In this and following figures, distances indicated on the axes are shown in parsec.}\label{LabelFig_JetLow2DZoom}
\end{center}
\end{figure*}

Further, an estimated propagation speed of the head of the 
jet can be obtained from using the expression for
pressure-matched jet propagation in 1D \citep{Martietal97, Rosenetal99}. 
For a cold external medium, this yields
\begin{equation}\label{headjet}
v_{\rm jet}^{1D}\,=\,\frac{\sqrt{\eta_{\rm R}}}{\sqrt{\eta_{\rm R}}+1}\,v_{\rm b}\,,
\end{equation}
where $\eta_{\rm R}=\gamma_{\rm b}^2\frac{\rho_{\rm b}h_{\rm b}}{\rho_{\rm m}h_{\rm m}}$ is the ratio 
between the inertia in the jet and in the external medium. In the relativistic
case, the inertia of the flow increases as $\gamma^2$ with the speed 
(which makes relativistic jets more stable than jets in young stellar objects). 
Our $t=0$ conditions then use the number density of the cold external medium in the ${\rm pc}$-scale region 
set to $n_{\rm Low}=1{\rm cm^{-3}}$ as a scaling value, together with $R_{\rm b}$ as a unit of length, in combination with
$c=1$. Then, the pressure in this cold external medium is set to $p=10^{-3}m_p\, n_{\rm Low} c^2$, 
and this value for the pressure is in fact taken equal in the jet, and also in the outer
region which is in static equilibrium with the inner region.
We take the beam Lorentz factor fixed at $\gamma_{\rm b}=20$ for the 
simulations A, B, E, F, J and at $\gamma_{\rm b}=10$ for the simulations 
C, D, G, H, I. These assumed high inlet Lorentz factors are at the observed 
values appropriate for FR II and BL Lac objects at ${\rm pc}$-scale \citep{Kellermannetal04, Cohenetal07}.
This is consistent with the hypothesis that BL Lac are FR I radio galaxies 
observed with a small angle to the line of sight \citep{Urry&Padovani95}. 

The density in the jet, and in principle also the pressure, can be deduced 
using Eq.~\ref{KLflux} 
and Eq.~\ref{headjet}, by imposing the jet head propagation Lorentz factor 
$\gamma_{\rm head, Low}$ in the low density medium
(in practice, this is limited by the condition to have a positive jet pressure 
and density). 
If we choose a value $\gamma_{\rm head, Low}=5$ ($v_{\rm head, Low} = 0.979796$),
we find from Eq.~\ref{headjet} a ratio between the jet beam inertia and the 
lower medium inertia. This is in the range $\eta_{\rm R, Low} = [2672.3,3000]$
 for our models with high 
energy  (A, B, G, H, I, J), while we have $\eta_{\rm R, Low} = [1.8, 2.7]$ for 
our models with lower energy (C, D, E, F). 
We anticipate from these inertia contrasts that the high energy
jet in the lower region is fairly stable and will
conserve its narrow structure, while the lower energy jets will already be disturbed by the external medium in the lower part.
 When we use Eq.~\ref{KLflux} to obtain our computational value for the density ratio between the jet beam and the external
medium in the lower scale, we find ${n_{\rm b}}/{n_{\rm Low}} = 6.681$ for the simulations A, B, J; 
${n_{\rm b}}/{n_{\rm Low}} = 29.99$ for the simulations G, H, I; ${n_{\rm b}}/{n_{\rm Low}} = 2.738\times 10^{-2}$ for the simulations 
C, D;  and finally ${n_{\rm b}}/{n_{\rm Low}} = 6.82\times 10^{-3}$ for the 
simulations E, F. 

For the density of the external medium in the upper, downstream region, we can argue similarly 
by setting a head propagation Lorentz factor at ${\rm kpc}$-scale.
In this paper, we will investigate many cases for the upper medium conditions.
In the two first cases (A, B), we choose a uniform density medium.
In all other models, beyond the jump in the density at the
interface between the lower and higher scale region, 
the density decreases with distance from the source with 
a simple power-law $Z_{\rm jump}/\sqrt{R^2+Z^2}$ \citep{Kaiser&Alexander97}.
Furthermore, in model A, we assume that the jet undergoes weak deceleration in this 
upper region, where the
Lorentz factor of the jet head drops to $\gamma_{\rm head, Up}=1.5$ 
(mildly relativistic). 
Then, the ratio between the jet beam inertia and upper medium inertia is $\eta_{\rm R, Up}=8.65$, 
which increases the influence of the external medium on the jet. 
Again, similar reasoning makes it plausible to use a density ratio
$n_{\rm Up}/n_{\rm jet}\sim 4.687$ for the upper medium then. 
In model B, we consider a very dense upper medium, expected to induce a strong
deceleration of the jet to  $\gamma_{\rm head, Up}=1.02$ 
($v_{\rm head, Up} = 0.197$), which corresponds to a sub-relativistic jet. 
The inertia ratio between the jet beam and the upper medium is in this case 
$\eta_{\rm R, Up}\sim 0.0604$, and this very low value will lead to increase
the jet-external medium interaction, and the growth of body and surface mode
instabilities in the jet.  
We find that to end up with this low Lorentz factor in the upper high density medium, we need to assume
a fairly extreme, high value of density $n_{\rm Up}/n_{\rm jet} \sim 671.22$. 
This is admittedly very high, however, many numerical simulations show that to decelerate a relativistic jet to 
sub-relativistic speeds, we need overdense external medium where $n_{\rm Up}/n_{\rm b}>100$ and even higher \citep{Krause05}. 
This is also seen by the exploited valued for the density contrast in most numerical simulations to date, as collected in
Table~\ref{table:1}.
Furthermore, from Eq.~\ref{KLflux} one can argue that for fixed luminosity,
the jet density decreases with jet radius as $(0.05 {\rm pc}/R_{\rm b})^2$.
This suggests that for bigger opening angle of the jet (hence bigger jet radius), the jet 
density and thus also the upper external medium density will be lower, while the 
behaviour of the jet in this region should be the same. 
For all other models the density ratio at the interface reaches values as indicated in Table~\ref{table:1}, and then decreases outwards.
To model also the effect of conical versus cylindrical jet propagation, we assume a $1^\circ$ opening angle at the inlet for models
D, F, H, I and J. This opening angle is taken smaller than the open angle of the jet at the boundary 
$3^\circ$, since the jet is supposed to collimate slowly during its propagation.

\subsection{Employed resolution}
In all our simulations of jet propagation, 
we set the lower boundary at
$Z_{\rm in}=0.5 \left(\frac{R_{\rm b}}{0.05 {\rm pc}}\right){\rm pc}$ and initialize the jet material
within the domain for radii $R< R_{\rm b}$ and 
extended to $Z=1.0 \left(\frac{R_{\rm b}}{0.05 {\rm pc}}\right){\rm pc}$.
The boundary imposes a stationary mass flux at the lower boundary for radii $R<R_{\rm b}$. The
lower boundary for $R>R_{\rm b}$ is open. The top boundary 
is set at $Z_{\rm ext}$ and the jump in density is set at a distance $Z_{\rm jump}=Z_{\rm ext}/2$, midway in our (very large) computational domain.
Note again that, as evident from Table~\ref{table:1}, we here cover jet propagation over distances that go as far as 400 jet beam radii 
and with enough resolution to analyse with high accuracy the interaction of the
jet with the external medium, 
and this is only feasible thanks to our AMR capabilities.

Simulation A  is done on a domain with size 
$\left[R,Z\right] \in \left[0,10\right]\times \left[10, 400\right]$ 
(in units of $R_{\rm b}$), with a resolution on the base level of $40\times 1200$. Our grid-adaptive runs
allow for 5 levels, achieving an effective resolution of $640 \times 19200$. 
The second simulation is done on a domain  with size 
$\left[R,Z\right] \in \left[0,40\right]\times \left[10, 400\right]$ 
and base level resolution of $160\times 1200$, however, we allow for 6
 levels achieving an effective resolution of $57 \times 38400$. 
This anticipates that since
the ratio between the jet beam density and upper medium density is very high in the second case,
the shock in this model will be very strong, and is likely to produce a turbulent cocoon and 
jet in this region. 
All other cases are
done on a domain with size $\left[R,Z\right] \in \left[0,40\right]\times \left[10, 400\right]$  and with an effective resolution $3072\times 4800$ (4 levels).
All the simulations exploit the hybrid version of HLLC, as explained in the appendix.

\section{Discussion of results}\label{secres}

From a basic point of view, the jet-external medium interaction is 
structured along the jet propagation axis in a way which is similar to the 1D
tests described in our appendix. At the head of the jet, there are the 
forward shock, the contact discontinuity (called the work surface), and the 
reverse shock (called the Mach disk). The forward shock compresses and heats 
the external medium, which spreads laterally, leading to the formation of a 
bow shock. The reverse shock decelerates the beam matter by converting its 
kinetic energy to thermal energy. However, the shapes of the Mach disk and 
forward shock in a true 2D jet are oblique, and other 2D effects appear which we 
describe in the following.

\subsection{Propagation through the uniform lower region}
\subsubsection{Models A, B and J}
In  models A, B, and J, the properties of the jet and of the lower external 
medium are the same, and the following discussion is applicable to these 3 cases.
The jet density, prior to penetrating the dense upper region, is seen in 
a close-up view in Fig.~\ref{LabelFig_JetLow2DZoom}.
This figure is at time $t=160$ (in units of light crossing time for the jet beam radius).
In the inner low density medium, the forward shock is relativistic. 
The jet also interacts laterally with the external medium, as there is a 
boundary shear layer. Due to the favorable density contrast, this layer does 
not disturb the jet propagation in this lower region. This thin shear layer is 
created as the Mach disk compresses the external shell of the jet. 
Only near the jet head, in between working surface and reverse shock (Mach disk), and
a bit beyond the reverse shock location, is the shear layer Kelvin-Helmholtz
unstable, see our (zoomed-in) Fig.~\ref{LabelFig_JetLow2DZoom}.
We find hardly any backflow from the work surface in this lower region.

\begin{figure}
\begin{center}
\FIG{
{\rotatebox{0}{\resizebox{0.48\columnwidth}{3.5cm}{\includegraphics{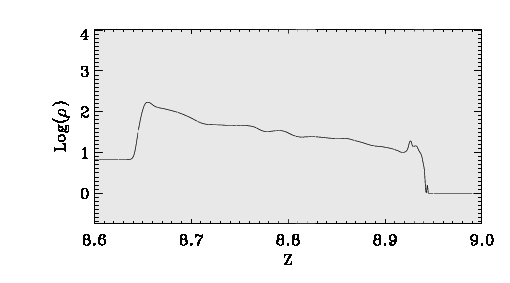}}}}
{\rotatebox{0}{\resizebox{0.48\columnwidth}{3.5cm}{\includegraphics{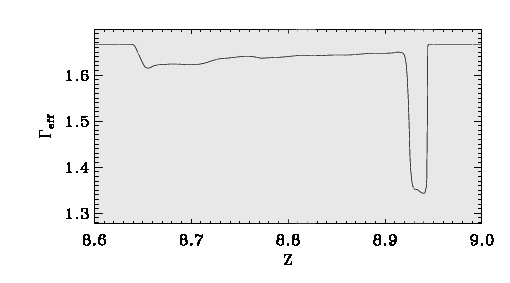}}}}
{\rotatebox{0}{\resizebox{0.48\columnwidth}{3.5cm}{\includegraphics{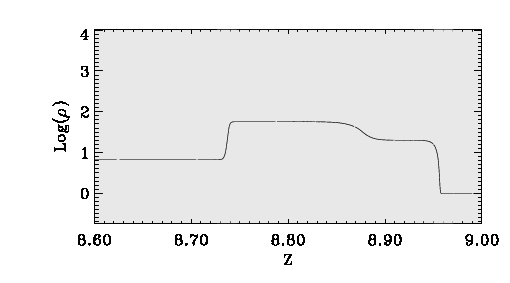}}}}
{\rotatebox{0}{\resizebox{0.48\columnwidth}{3.5cm}{\includegraphics{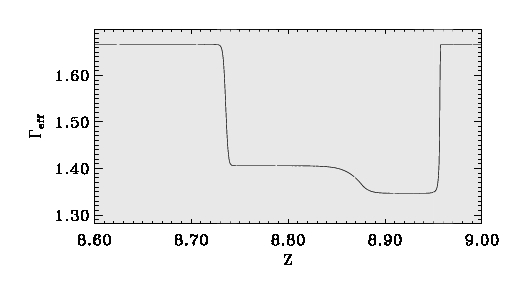}}}}
}
\caption{Top: A cut along the Z axis through the jet head at $t=160$, prior to penetrating the denser medium (cases A, B and J). Shown is: (Left) the logarithm of density,
(Right) the effective polytropic index. Bottom: a 1D equivalent relativistic shock tube problem.}\label{LabelFig_JetLow1DZoom}
\end{center}
\end{figure}

Focusing on the axial structure of the jet head, 
Fig.~\ref{LabelFig_JetLow1DZoom} compares the obtained on-axis structure with a first equivalent 1D
Riemann problem. 
The sound speed in the shocked external medium (between forward shock and contact or work surface)
increases to reach the maximum value allowed by the equation of state 
$c_{s}\sim 0.636$, hence the effective polytropic index in this flow drops to 
$\Gamma_{\rm eff}\sim 1.34$ (a near ultra-relativistic state). 
Then the compression is very high and 
hence the distance between the contact discontinuity and forward shock remains 
small: we obtain a spacing of about $\delta X=0.044$ pc at $ t=160$. 
This is even three times smaller than that obtained in an equivalent 1D shock problem, as seen in
Fig.~\ref{LabelFig_JetLow1DZoom}. The difference is induced by the lateral 
spreading of the shocked shell of swept-up matter, which acts to decrease the 
amount of accreted mass. 
Indeed, shocked material in front of the beam will start to spread laterally, due 
to the strong (radial) gradient in the pressure and in the inertia 
between the shocked material and the cold, static external medium. This initial 
spreading can even be quantified from a second 1D Riemann problem in the 
direction across the beam (which is not shown here), 
and this confirms the 2D result which shows an 
initial supersonic lateral speed $v_{\rm lateral}\sim 0.7973$ for the shocked material. 
This outwards spreading leads to the formation of a bow shock, whose lateral
spreading eventually (a 2D effect) decelerates to $v_{\rm lateral}\sim 0.36$. 
With this fast lateral spreading of the shocked material, no complex cocoon 
forms in this region of low density for these three cases A, B and J. 
The second equivalent Riemann problem in the radial direction, also shows how a rarefaction wave 
propagates laterally inward into the swept-up matter by the jet.
This rarefaction wave moves slowly towards the axis at approximate speed $v_{\rm r}=0.058$, 
and matter accumulates mainly near the jet axis. This produces 
relatively more deceleration of the jet near the axis, and hence to a radially structured jet head.

\begin{figure*}
\begin{center}
\FIG{
{\rotatebox{90}{\resizebox{5cm}{\textwidth}{\includegraphics{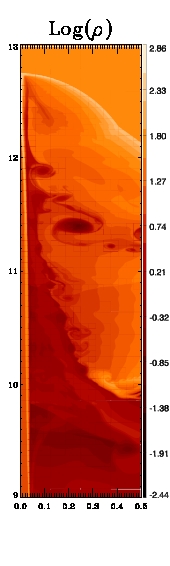}}}}
{\rotatebox{90}{\resizebox{5cm}{\textwidth}{\includegraphics{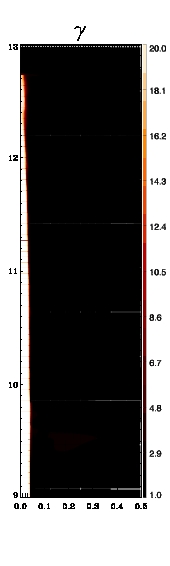}}}}
{\rotatebox{90}{\resizebox{5cm}{\textwidth}{\includegraphics{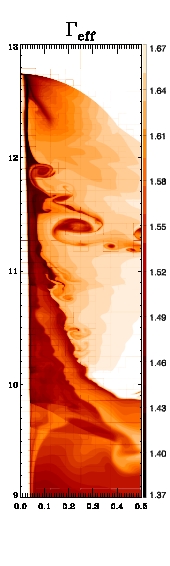}}}}
}
\caption{A zoom on the jet head after penetrating the denser medium, for case A. We zoom with $R,Z \in 
\left[0,0.5\right]\times\left[9,13\right]$ parsec, at $t=240$. Top: logarithm of density, middle: Lorentz factor, bottom: 
effective polytropic index.}\label{LabelFig_Jet2DUp}
\end{center}
\end{figure*}

Returning to the axial structure as shown in Fig.~\ref{LabelFig_JetLow1DZoom},
in this lower density region at $t=160$, 
the reverse shock (or Mach disk) is Newtonian with
$\gamma_{\rm rs}\sim 1.013$. This in fact, is totally different from the equivalent 1D
problem, as clearly seen in effective polytropic index in Fig.~\ref{LabelFig_JetLow1DZoom}. 
The Mach disk compresses the beam matter, increasing the density 
and decelerating the flow to $\gamma\sim 18$ 
(in the equivalent 1D case the shocked jet beam should have $\gamma=5$). 
Furthermore, the 2D result shows a relatively larger region of the jet head which is formed by shocked beam matter (compare the 1D cuts in Fig.~\ref{LabelFig_JetLow1DZoom}).
The sound speed of the shocked beam material reaches $\sim 0.2$ with an 
effective polytropic index $\sim 1.64$. Also this shocked beam material tends to expand laterally, 
with a speed $v_{\rm lateral}\sim  0.1$. 
Similar to the lateral dynamics of the swept-up, shocked matter, 
again a rarefaction wave propagates towards the axis, which progresses very slowly,
and eventually the shocked jet head ends up denser and colder.
Since the shocked external medium (in the forward bow shock) has already cleared the immediate 
surroundings of the beam, the shocked beam matter will hardly form a
backflow. 

The remarkable feature of jet beam propagation in a low density region is 
thus its stability. There is a very weak influence of the external medium on the jet, which 
allows the jet to transport energy over a long distance, and we followed this 
process over about 200 light crossing times of the jet beam radius.
Only a small fraction of the jet energy is 
transferred to the external medium in the inner region, through the bow shock.
During the propagation in the inner region, there is no strong backflow and we hardly see
the formation of a turbulent shear layer. 
In this lower region, the Lorentz factor of the jet beam, and even the shocked
jet beam, remains relativistic with $\gamma\sim 20$.
In turn, the jet head propagates relativistically 
with a Lorentz factor $5$. It must be clear from the above discussion
that even in the lower density region, the inclusion of a realistic EOS is necessary as the
forward shock is relativistic and the reverse shock (Mach disk) is Newtonian.

\subsubsection{Other models when traversing the lower medium}

In model C, which is a light jet in this lower region at beam Lorentz factor 10, the jet interacts with the external medium mainly through the bow shock. 
Due to its lighter density, a turbulent cocoon gets formed during propagation.
The turbulent cocoon interacts also laterally with the jet beam by compressing it, inducing internal shocks in the jet beam. The distance between shocks 
is about $\lambda_{\rm shock}=20 R_{\rm b}$. 
At each shock the jet undergoes a small deceleration to Lorentz factor $\gamma\sim 9$ and is accelerated again to Lorentz factor $\gamma\sim 10$ due to the 
expansion of the jet after each shock.
Thus, in this lower region the jet remains relativistic, its energy is mainly kinetic, and only a very weak fraction of the energy is transferred to the external medium by entrainment. 
The interaction of this light jet with the external medium also shows the formation of a low speed, higher pressure sheet of thickness $\sim R_{\rm b}/2$  surrounding the inner jet.

Model D is similar to C, but has the jet emerging with an initial open angle of $\theta = 1^{\circ}$.
In the first part the D jet undergoes a free expansion, and its density and pressure decrease until the pressure in the surrounding cocoon becomes high enough to compensate the jet thermal and ram pressure in the lateral direction. One can witness that in this region, an oblique shock forms, which moves with a slow speed $v\sim 0.3$ that decreases during the jet propagation. 
In fact, as the jet propagates forward, the pressure of the cocoon behind the jet head decreases with cocoon expansion, and then the oblique shock moves forward. At $t=820$, the oblique shock is at normalise distance $Z\sim 5$ Fig.(\ref{LabelFig_Jet2DUpHighZ2B})
This oblique shock in fact decelerates the jet to $\gamma_{\rm b}\sim 8$ and consecutively collimates it cylindrically, to end up with a smaller radius than 
the initial one imposed at inlet. Beyond the oblique shock, the jet beam is unstable and subject to the development of internal shocks. The amount of the matter entrained by the jet is larger than in the C case (cylindrical inlet). In this part of the jet, the internal energy increases and the state of the matter becomes mildly relativistic with an effective polytropic index falling to $\Gamma_{\rm eff} \sim 1.55$.
The jet in this model D is then more stable than the jet in the model C, since the initial conical propagation of the jet decreases the lateral interaction 
of the jet with the external medium and then the jet propagates in a laminar way until it reaches the collimation shock.

The model E differs from C only in its twice higher beam Lorentz factor. It has an inertia ratio between jet-external medium of similar order than in 
model C. Despite the faster jet beam, its Lorentz factor along the beam in the first region ends up oscillating between $10-25$ due to the successive internal shocks and jet expansions, such that the jets in both model have the same general behavior.
In the model F, the E (cylindrical) jet model now emerges conically with an initial open angle of $\theta = 1^{\circ}$.
Its Lorentz factor $\gamma_{\rm b}=20$ and kinetic luminosity 
$L_{\rm Jet, Kin}=10^{43} {\rm ergs/s}$, makes that the difference with the conical model D is that the jet is faster. In this model F, the forward shock 
that forms at the jet head is stronger producing higher pressure and a more extended cocoon. This cocoon limits the region of the free conical expansion of 
the jet by forming oblique shocks that recollimate the jet. This oblique shock in model F propagates with a speed 
$v\sim 0.1$, which is lower than the speed of the oblique shock in the model D. As in model D, the speed of the oblique 
shock decreases during the propagation. Beyond this oblique shock, the Lorentz factor of the jet decreases to about $10$, and the jet starts to be unstable 
with the formation of internal shocks that compress and induce consecutive acceleration and deceleration of the jet. In this region the state 
of the matter in the jet is relativistic with an effective polytropic index $\Gamma_{\rm eff}\sim 1.45$.  At $t=820$, the oblique shock is at normalise distance $Z\sim 7.5$ Fig.(\ref{LabelFig_Jet2DUpHighZ2B})
 
In model G, the interaction between the jet and the external medium
is roughly the same than in model A, because only the Lorentz factor of the jet beam changes from $\gamma=20$ in model A to $\gamma=10$.
In fact, the main differences between the two models is in the details of the structure of the jet head, since the rate of compression at the front shock 
and the Mach disk (reverse shock) is different.  In any case, the jet interacts weakly laterally with the external medium and it remains stable, consistent with its higher density with respect to the surroundings.
Finally, in the conical models H and I (otherwise similar to G and also of high energy), the jet interacts with external medium only through the shock in front and there is no 
lateral interaction between the jet and the external medium, as only the head of the jet is a bit collimated. In fact, the conical propagation of the jet 
decreases the influence of the low density external medium on the jet.

\subsubsection{Summary for lower region}
In the lower region, the jet head propagates with a speed  $v_{head, Low}\sim 0.6$ for the models C, D, E, F, and with a speed $v_{head, Low}s\sim 0.99$ in 
the models A, B, G, H, I, J, as collected also in Fig.~\ref{JetHeadCompart}. In fact, these two groups are clearly 
distinct from the (input) kinetic luminosity of the jet beam. The low energy first group
finds the lighter density jets interacting strongly with the external medium and slowing down, since in this group the
power of the jet beam is $L_{\rm jet,Kin} \sim 10^{43} \,{\rm ergs/s}$ and then
the inertia ratio between the jet and the external medium is low, and at most of order $\eta_{\rm R}\sim 2.73$.  
For the second high energy group, the external medium influences weakly the jet, since the power of the jet beam is 
$L_{\rm jet,Kin} \sim 10^{46} \,{\rm ergs/s}$ and then the inertia ratio between the jet and the external medium is high, up to order $\eta_{\rm R}\sim 3000$.
Thus the variation in the Lorentz factor $10$ to $20$ of the jet beam and the small variation in the opening angle of the jet (cylindrical case versus
conical with a small opening angle $1^{\circ}$) do not significantly influence the speed of propagation of the jet head in the lower scale region, which
is here simulated up to $200$ jet radii.
While we pointed out various aspects that are clearly captured only using a relatistic
EOS model, our findings for jet propagation through the uniform, lower region are fully consistent 
with earlier works.

\subsection{Upper region propagation}

\subsubsection{Model A}
We now turn to the second stage in the dynamics, after the jet has passed the density jump.
In the model  A, the density ratio between the jet and upper external
medium suddenly changes to $\rho_{\rm Up}/\rho_{\rm b} \sim 4.687$, making it a light jet.
In this denser upper region, the initial interaction between jet and 
dense upper medium is shown in a zoomed Fig.~\ref{LabelFig_Jet2DUp} at 
time $t=240$. First, as the jet penetrates 
the denser medium, an oblique shock propagates laterally in the upper medium. 
The relatively higher density of the upper medium gives rise to reflection of this shocked matter and hence 
produces a somewhat more turbulent and hot cocoon. At the jet surface, a thin, rarefied, and 
very hot region develops (seen best in the low effective polytropic index in 
Fig.~\ref{LabelFig_Jet2DUp}). 
In fact, the head of the jet now sweeps up more matter, enhancing the temperature.
The high pressure in this region in turn gives rise to another oblique 
shock, which is weaker but propagates upstream and toward the axis. This confines the 
jet. This shock is located at time $t=240$ at about $Z\sim 9.8$. 
The jet radius drops due to this compression to about $R=0.03$.

\begin{figure*}
\begin{center}
\FIG{
{\rotatebox{0}{\resizebox{5.0cm}{3.5cm}{\includegraphics{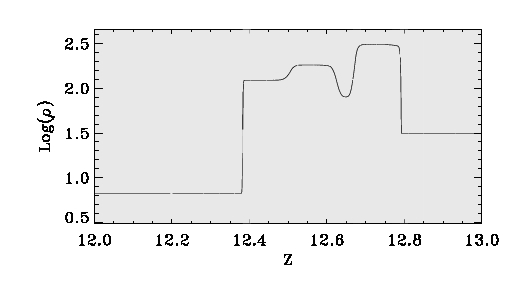}}}}
{\rotatebox{0}{\resizebox{5.0cm}{3.5cm}{\includegraphics{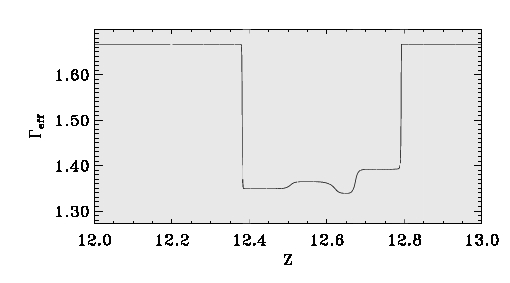}}}}
{\rotatebox{0}{\resizebox{5.0cm}{3.5cm}{\includegraphics{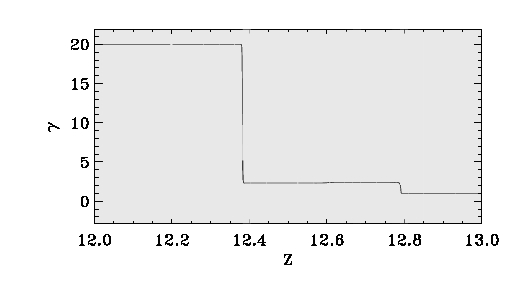}}}}
{\rotatebox{0}{\resizebox{5.0cm}{3.5cm}{\includegraphics{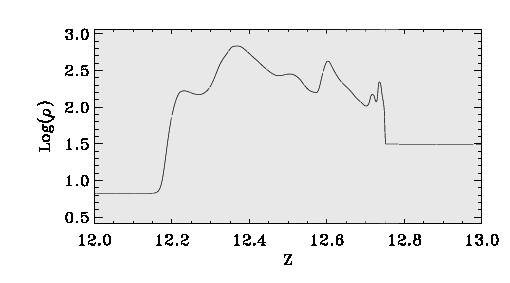}}}}
{\rotatebox{0}{\resizebox{5.0cm}{3.5cm}{\includegraphics{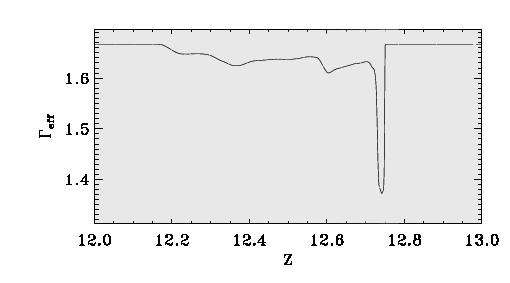}}}}
{\rotatebox{0}{\resizebox{5.0cm}{3.5cm}{\includegraphics{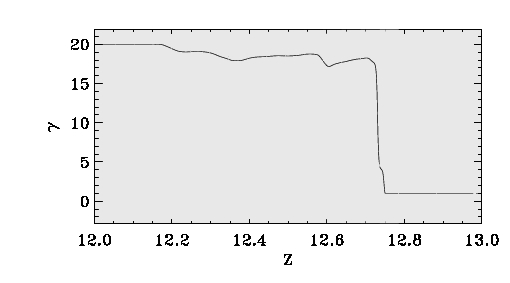}}}}
{\rotatebox{0}{\resizebox{5.0cm}{3.5cm}{\includegraphics{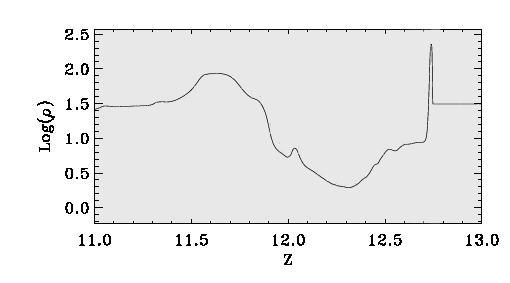}}}}
{\rotatebox{0}{\resizebox{5.0cm}{3.5cm}{\includegraphics{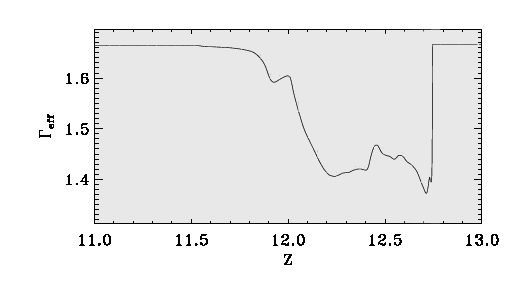}}}}
{\rotatebox{0}{\resizebox{5.0cm}{3.5cm}{\includegraphics{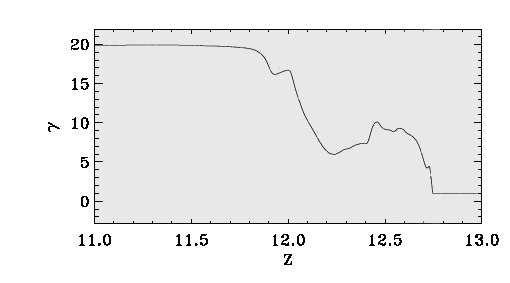}}}}
}
\caption{To analyze case A in the upper medium: we show at Left: logarithm of density, center: effective polytropic index, right: the Lorentz factor, at time 
$t=240$, i.e. after penetrating the denser medium. The three rows correspond to: (Top) a 1D equivalent relativistic shock problem, (middle) a cut along the Z axis 
(i.e. $R=0$) through
the jet head, and (Bottom) a cut along the axis at a radius $R=0.02$. 
From the top to the bottom, note the scales difference: $Z\in \left[12.0, 13.0\right]$ for top and middle, while
bottom panel has $Z\in \left[11.0, 13.0\right]$.}\label{LabelFig_JetUp1DZoom}
\end{center}
\end{figure*}

\begin{figure*}
\begin{center}
\FIG{
{\rotatebox{90}{\resizebox{5cm}{\textwidth}{\includegraphics{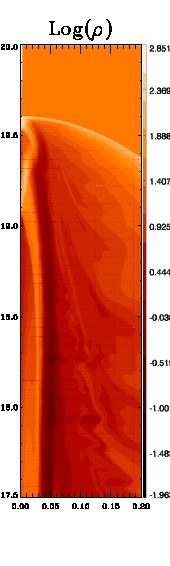}}}}
{\rotatebox{90}{\resizebox{5cm}{\textwidth}{\includegraphics{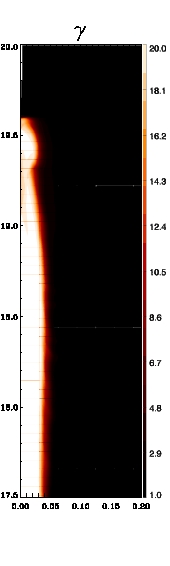}}}}
}
\caption{A zoom on the jet head with $R,Z \in
\left[0,0.2\right]\times\left[17.5,20\right]$, at $t=380$, for case A. (Top) Logarithm of density, (middle) Lorentz factor, 
(bottom) effective polytropic index.}\label{LabelFig_Jet2DUp2}
\end{center}
\end{figure*}

Again we can learn from the analogous 1D Riemann problem and compare it with the axial structure of the
2D result. This is done in Fig.~\ref{LabelFig_JetUp1DZoom}, at a time corresponding to
Fig.~\ref{LabelFig_Jet2DUp}.
As a 1D shock-structured jet head penetrates in high density medium in an upper region, there are 
in fact four layers of shocked material that can be distinguished along the 
beam axis. Once the previously formed forward (bow) shock meets up with the 
density jump, a new forward shock develops which seperates swept-up high 
density external medium from static external medium. A second contact or work 
surface from then on seperates shocked high density matter with previously 
swept-up, shocked, lower density matter. From this same location, a new 
reverse shock forms, which traverses the previously formed structured 
jet head (i.e. consisting of shocked lower matter, shocked beam matter, and 
ending with the old Mach disk or reverse shock). These add up to 5 discontinuities in total,
which are clearly seen in the analogous 1D problem shown in Fig.~\ref{LabelFig_JetUp1DZoom}.
However, the 2D jet propagation is rather different from this
1D model. The difference is quantified in Fig.~\ref{LabelFig_JetUp1DZoom}, 
as we draw the cut along the $Z$ axis (middle row), and also at a fixed
radial distance away from the axis at $R=0.02$ (bottom). The latter radius still
remains in the jet `spine', since the laterally bounding shear region which we also can detect in our jet beam variation extends from 
$0.025$ to $0.03$ parsec. 

We can clearly see that, when comparing these three cases, in 2D we find faster beam
flows immediately behind the front shock, which are also relativistic with a Lorentz 
factor $\gamma\sim 4$. This is accompanied by strong compression, as 
the effective polytropic index drops to $\Gamma_{\rm eff}\sim 1.37$. This is as in the 1D case. 
To be precise, at time $t=240$, the front shock reaches in the 2D case a distance $Z\sim  12.75 {\rm pc}$, where
the 1D reaches $Z \sim 12.8$. Behind the front shock in the 2D case, various internal shocks develop, partly
induced by the initial structure of the jet beam with its interaction with
the denser medium. At each new oblique shock, the flow undergoes a weak acceleration behind it.
In the region of the shocked beam, the Lorentz factor along the axis is 
oscillating between $\gamma_{\rm min}\sim 16$ and  
$\gamma_{\rm max}\sim 18$. Now, the trailing reverse shock (Mach disk) and other 
shocks in front of it are Newtonian along the axis, and the effective polytropic index oscillates 
between $\Gamma_{\rm eff, min} \sim 1.65$ and $\Gamma_{\rm eff, max} \sim 1.61$. 
In contrast, along the spine at fixed radius $R=0.02$, the reverse shock (Mach disk) and all 
shocks in front of it are stronger. In fact, the Mach disk is near-Newtonian (effective
polytropic index $\Gamma_{\rm eff}\sim 1.58$), whereas the other shocks are 
relativistic with effective polytropic index oscillating between 
$\gamma_{\rm min}\sim 1.34$ and $\gamma_{\rm max}\sim 1.46$. 

\begin{figure}
\begin{center}
\FIG{
{\rotatebox{0}{\resizebox{\columnwidth}{3.5cm}{\includegraphics{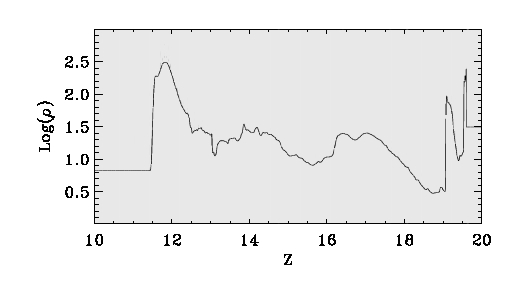}}}}
{\rotatebox{0}{\resizebox{\columnwidth}{3.5cm}{\includegraphics{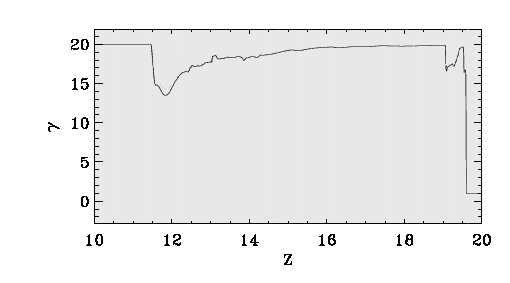}}}}
}
\caption{A cut through the jet for case A along the $Z$ axis, 
at $t=380$. (Top) the logarithm of density, (Bottom) the Lorentz factor.}\label{LabelFig_Jet1DUp2}
\end{center}
\end{figure}

There are thus clear 2D effects, such as the
lateral compression of the jet by the hot cocoon, and
the presence of the oblique Mach disk. The latter at the same time shocks 
the jet beam, while compressing it laterally.
A kind of reflection happens (see also Fig.~\ref{LabelFig_Jet2DUp}), and 
in front of the reflection point, the jet radius increases again, and gives rise to a weak acceleration. 
We find almost no backflow along the jet in this upper region. In fact, parallel to the jet, the
hot shell of shocked external medium (as discussed above), moves
in the same direction with a speed $v_{\rm z}\sim 0.4$. The difference of speed
between this shell and the surrounding slower and denser cocoon, makes
this region subject to instability, as seen clearly in Fig.~\ref{LabelFig_Jet2DUp}. Meanwhile,
the jet itself remains relatively stable. It seems that the formation of dense shear in the outer 
region of the jet increases the jet stability. However, in the jet,
weak internal shocks are induced by the compression. The end
result is the formation of a knot with long wavelength, and this result is 
shown in Fig.~\ref{LabelFig_Jet2DUp2} where we plot the logarithm of the 
density and Lorentz factor at a later time
$t=380$. We also show the variation of density and Lorentz 
factor along a larger section of the jet axis at this endtime, in Fig.~\ref{LabelFig_Jet1DUp2}.
It can be seen that several internal shocks have developed, reminiscent of knots.

From this first model, it appears that a mildly overdense external medium 
can not induce the strong deceleration observed in FR I jets, as the jet 
head for this case A remains relativistic in this region. However, this model could be
relevant for FR II jets, as it reproduces the weak interaction with the external 
medium in the inner region, and stays stably structured in the upper region.

\subsubsection{Upper region: model B}

\begin{figure}
\begin{center}
\FIG{
{\rotatebox{90}{\resizebox{20cm}{0.45\textwidth}{\includegraphics{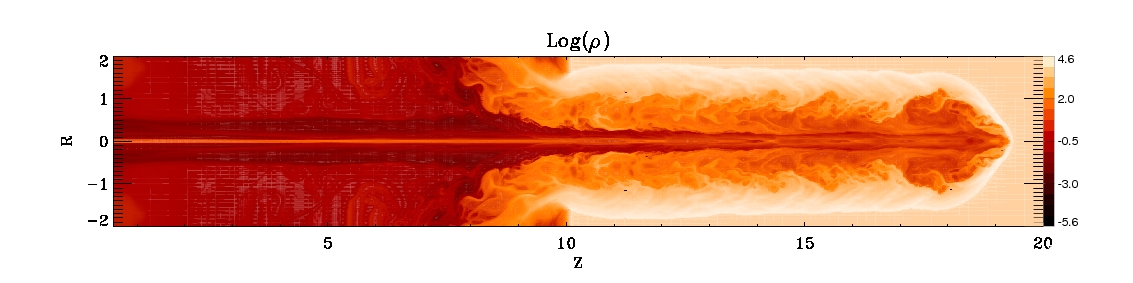}}}}
}
\caption{The logarithm of density at $t=900$ on the entire simulation domain, for the high density upper medium case B.}\label{LabelFig_Jet2DUpHigh15}
\end{center}
\end{figure}

\begin{figure*}
\begin{center}
\FIG{
{\rotatebox{90}{\resizebox{5cm}{\textwidth}{\includegraphics{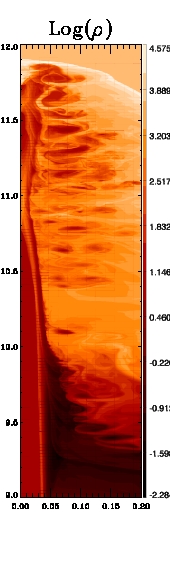}}}}
{\rotatebox{90}{\resizebox{5cm}{\textwidth}{\includegraphics{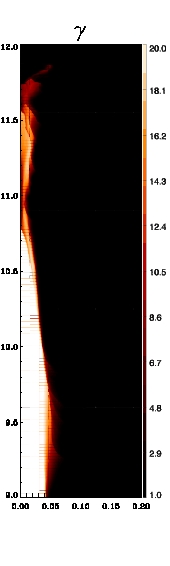}}}}
{\rotatebox{90}{\resizebox{5cm}{\textwidth}{\includegraphics{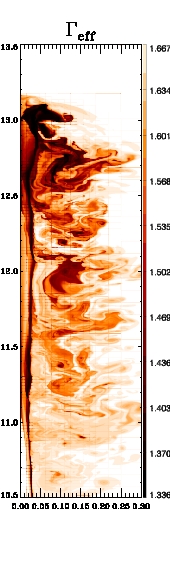}}}}
}
\caption{A zoom on the jet head after penetrating the high density medium in case B, for $R,Z \in \left[0,0.2\right]\times\left[9,12\right]$ Top and middle panel show at $t=240$: (Top) Logarithm of density, (middle) Lorentz factor. Bottom panel shows, later at $t=300$, the effective polytropic index,
zoomed with $R,Z \in \left[0,0.3\right]\times\left[10.5,13.5\right]$.
}\label{LabelFig_Jet2DUpHigh}
\end{center}
\end{figure*}

In the model B, the density ratio between the jet and the denser external
medium is increased to $\rho_{\rm Up}/\rho_{\rm b}= 671.220$.
The interaction of the jet with this very dense external medium is very strong
and entails the formation of a turbulent cocoon. In Fig.~\ref{LabelFig_Jet2DUpHigh15}, we show the
density in the entire simulation region, at our endtime $t=900$.
In this model, the inertia ratio between jet and external 
medium was small $\eta \sim 0.0604$, making the jet 
beam subject to instabilities. A zoom of the jet head at this same endtime is also shown in the last panel from
Fig.~\ref{LabelFig_Jet2DUpHigh}.

\begin{figure*}
\begin{center}
\FIG{
{\rotatebox{0}{\resizebox{5.0cm}{3.5cm}{\includegraphics{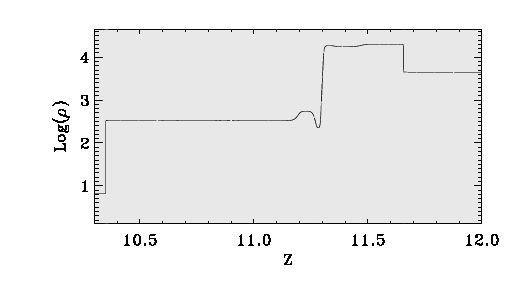}}}}
{\rotatebox{0}{\resizebox{5.0cm}{3.5cm}{\includegraphics{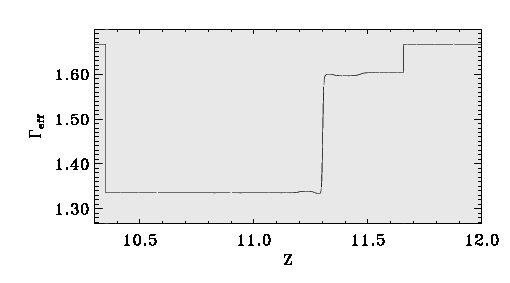}}}}
{\rotatebox{0}{\resizebox{5.0cm}{3.5cm}{\includegraphics{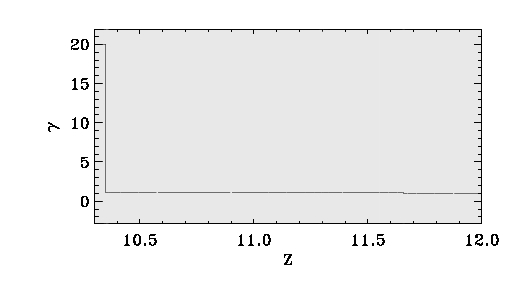}}}}
{\rotatebox{0}{\resizebox{5.0cm}{3.5cm}{\includegraphics{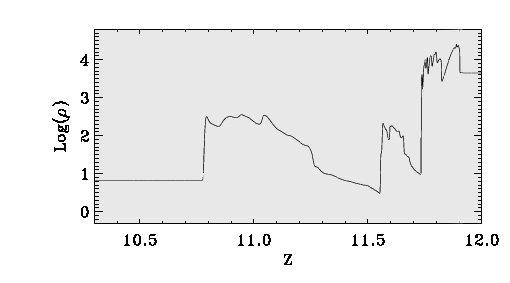}}}}
{\rotatebox{0}{\resizebox{5.0cm}{3.5cm}{\includegraphics{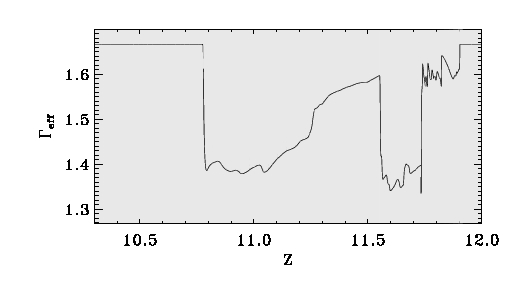}}}}
{\rotatebox{0}{\resizebox{5.0cm}{3.5cm}{\includegraphics{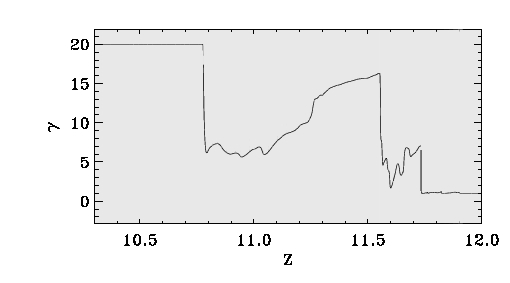}}}}
{\rotatebox{0}{\resizebox{5.0cm}{3.5cm}{\includegraphics{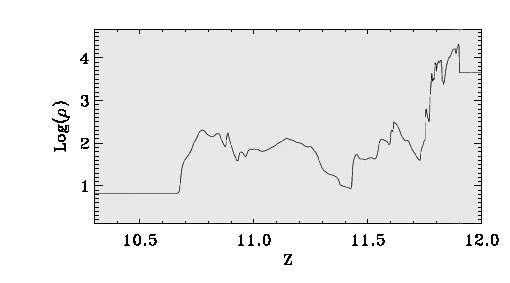}}}}
{\rotatebox{0}{\resizebox{5.0cm}{3.5cm}{\includegraphics{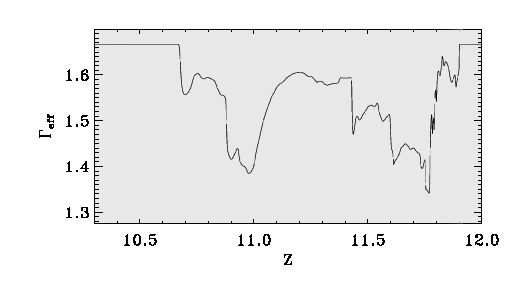}}}}
{\rotatebox{0}{\resizebox{5.0cm}{3.5cm}{\includegraphics{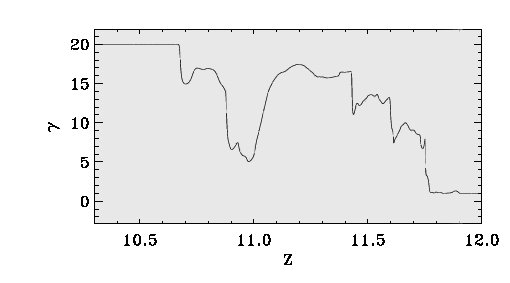}}}}
}
\caption{Shown are: (Left) logarithm of density, (center) effective polytropic index,
 (Right) Lorentz factor, for three cases. The three rows are
(Top) a 1D equivalent relativistic shock problem, (middle) a cut along the Z axis on the case B jet head at $t=240$. 
(bottom) A cut along the axis at fixed $R=0.01$.
}\label{LabelFig_JetUp1DZoom2}
\end{center}
\end{figure*}

The result of the interaction is the formation
of a bow shock propagating in front of the jet beam and to the side. The
shocked external medium initially spreads laterally with a speed
$v_{\rm lateral}\sim 0.18$, lower than the sound speed. The high 
density of the external medium leads to a weak compression rate, and
in fact, the front (forward) shock is now Newtonian with an 
effective polytropic index $\Gamma_{\rm eff}\sim 1.6$. The compression 
rate is about $5$ and the jet head propagates with a Lorentz factor less than $1.25$ along
the axis and more slowly away from the axis. Indeed, along the
line $R=0.01$ the shock propagates
with a Lorentz factor $1.09$ (Fig.~\ref{LabelFig_JetUp1DZoom2}). 

The slow spreading of shocked dense matter makes the 2D effects on the propagation 
of the forward shock less significant than in the first model, and there is less difference with a similar 1D case.
Again, Fig.~\ref{LabelFig_JetUp1DZoom2} compares
analogous 1D results with axial cuts along axis and at $R=0.01$, for an earlier time $t=240$ 
(for which 2D impressions are shown in the top two panels of Fig.~\ref{LabelFig_Jet2DUpHigh}).
The shocked jet beam in 2D is structured axially as in the 1D case with the appearance of a new work 
surface, and the Mach disk, while the compression rate at the trailing
Mach disk approaches the 1D case. However, we see that the effective polytropic index there is higher, namely
$\Gamma_{\rm eff}\sim 1.4$, and the Lorentz factor is different
($\gamma\sim 1.03$ in the 1D and $\gamma\sim 6$ in the 2D jet). 
The resemblance of the structure of the 
shocked external medium and the jet beam in 1D versus 2D is better
near the jet axis. We can clearly detect the new forward shock, and the 
second contact discontinuity, seperating the shocked high density matter with previously 
swept-up matter from the lower density region. 
The differences between 2D and 1D are the additional oscillations on the density and pressure,
which are due to the lateral effects. Sideways spreading matter is strongly reflected,
as the high density of the external medium slows down the lateral growth of the cocoon.
Moreover, the oblique Mach disk confines the jet. Beyond it, the jet radius
increases and the flow is accelerated again to reach a Lorentz factor 
$\gamma\sim 17$. Then, the hot and dense cocoon compresses the jet again, and
a new internal shock decelerates the jet. Along the
line $R=0.01$, the reverse shock is 
Newtonian and there is evidence of more internal shock development, indicating
complex interaction with the cocoon. 

In this case B, we do find strong backflows. The shocked jet beam matter by the trailing Mach disk 
again is subject to sideways spreading. The denser hot cocoon reflects this spreading matter,
and induces the formation of a backflow. In fact, between the jet and the shocked
external medium, a shear layer develops, consisting of a dense and relatively hot flow
propagating upstream with a speed of order $v_{\rm backflow}\sim -0.2$.
Near the top of the jet head, the backflow is essentialy consisting of 
shocked jet beam matter, flowing in a cylindrical shell of thickness $\Delta R\sim 
0.01$ parsec. 
The mass flux in this backflow increases in the upstream direction.
Moreover, this backflow entrains with it shocked external cocoon matter, and 
then carries more mass, which slows down its speed to $v_{\rm backflow}\sim -0.1$
when it reacheas $Z=10.0$ (the original position of the density jump in the external medium). 
The backflow interacts also strongly with the non-shocked jet beam, and
it compresses the jet beam, and induces shocks. The final result of this fairly
complex interaction is confinement and overall deceleration of the jet in the upper region. 
The jet radius decreases to 
$R\sim 0.02$ in $Z\sim 5.0$, at $t\sim 900.0$. Beyond this point the jet
radii starts to increase. And at the interface between the two medium a strong
oblique shock form. The jet is strongly decelerate at this shock to 
$\gamma\sim 2.0 $ 
and  the it pressure increases, such the relativistic mach number
fall to $M\sim 5$. The high pressure jet expand lateraly in this region 
and to radius $R\sim 6\times R_{\rm jet, b}$. 

The interaction of the backflow with the shocked external medium in the cocoon
induces the development of Kelvin-Helmholtz instabilities.
This is the result of the velocity gradient between the backflow and the cocoon
and also because the backflow is slightly underdense with respect to the cocoon.
It is clear from Fig.~\ref{LabelFig_Jet2DUpHigh} that
the entire region is highly structured due to instability development.
In this upper region, the density ratio was $\rho_{\rm Up}/\rho_{\rm b}= 
671.220$, and we indeed find a transition to a beam suddenly traveling at a 
Lorentz factor of $\gamma_{\rm Up} \sim 1.25$. 
This value is a bit higher than the value 
used for setting up the initial condition (where we used $1.02$).
This difference can be explained, since the head of the jet 
beam is made up of hot swept-up matter and hot shocked beam matter, while the 
estimate assumed cold conditions. Moreover, in this upper region, the forward 
bow shock is now Newtonian.
The various layers at the head of the jet give rise to strong turbulence 
development. 
As result of this interaction and the entrainment of the externa matter by
the jet is the deceleration of the jet in the upper
region to $\gamma\sim 1.5$ ($v\sim 0.3$).

When estimating the overall energy budget, we find that in the jet interaction with the upper medium, about of  $58\%$ jet 
energy gets deposited in the upper lobe, while a fraction of order $10\%$  is reflected in the lower region.

\subsubsection{Other cases: effects of opening angle and density decrease}

{\it Low energy jets.}\\

In the model C, the density ratio between the jet and the dense external medium is $\rho_{\rm b}/\rho_{\rm Up}=4.91\times 10^{-3}$. There is strong 
interaction of this light jet with the external medium, and this implies the formation of a high pressure cocoon. This turbulent cocoon disturbs the jet by 
inducing a strong backflow. The difference in pressure between turbulent cocoon and the colder lower region produces also backflows propagating into the lower region.
At the interface between the two regions, the jet beam is compressed. The jet becomes unstable and its pressure increases.
However, the jet remains relativistic in the upper region, since it decelerates only to a Lorentz factor $\gamma \sim 8$.
The decrease of the density in the external medium as assumed for this jet (see Table~\ref{table:1}) and the (minor) deceleration of the jet makes 
the forward shock weaker and then the pressure of the cocoon at the head of the jet decreases. Therefore, the jet beam radius actually increases slowly 
during propagation in this upper stratified region.

In the model D, the jet beam that finally emerges from the lower region is collimated cylindrically (despite its original opening angle, see the discussion in the previous paragraphs) and ends up with a relativistic Mach number $M \sim 20$ and with a Lorentz factor of $\gamma\sim 8$. This jet interacts again
strongly with the external, upper medium forming a high pressure cocoon that induces a backflow, also propagating upstream in the low density region. 
This backflow compresses the jet near the (perturbed) interface, increasing its density and pressure. The jet in the upper region is found to decelerate 
to a Lorentz factor $\gamma\sim 5$ through multiple internal shocks. As in model C, the compression of the jet at the interface increases its pressure and 
the assumed density decrease in the external medium causes the jet to undergo a slight conical expansion in this upper medium.

In the model E (main difference with C is that the jet is faster), we find an even stronger interaction with the external medium.
This configuration enhances the development of the shear instabilities which causes entrainment of ambient material. Then this jet E is found to decelerate smoothly to a Lorentz factor $\gamma \sim 5$, while it was characterized by $\gamma\sim 20$ on inlet.

In model F, the jet beam emerges from the upstream, lower region with a fairly low mach number $M \sim 15$ (lower than in model D) and with an internal energy of the order of the mass energy. Moreover, the jet in F propagates with a high Lorentz factor and higher thermal energy. When it interacts with the downstream, denser medium, it forms a flattened (compressed) and turbulent cocoon. Then a stronger backflow develops than in case D, that propagates upstream in the low density region. This backflow this time eventually induces a deceleration of the jet to a Lorentz factor $\gamma \sim 10$ (along the axis).
At the interface, the jet pressure ends up higher than the pressure of the cocoon, and also this jet expands in the upper region in a conical shape until it  reaches pressure equilibrium with the cocoon in this region. 
The jet interacts strongly laterally with the cocoon in this region, and overall, the jet beam slows to a Lorentz factor $\gamma\sim 5$ (a long the axis) through multiple internal shocks induced by instabilities and entrainment of external matter. \\

{\it High energy jets.}\\

\begin{figure*}
\begin{center}
\FIG{
{\rotatebox{90}{\resizebox{24cm}{0.23\textwidth}{\includegraphics{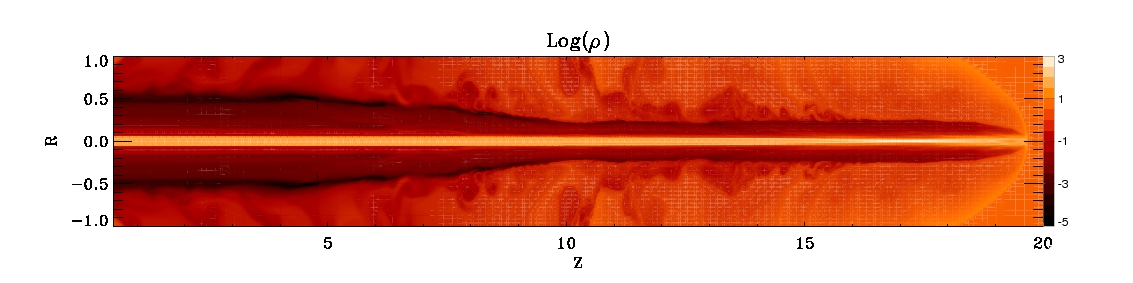}}}}
{\rotatebox{90}{\resizebox{24cm}{0.23\textwidth}{\includegraphics{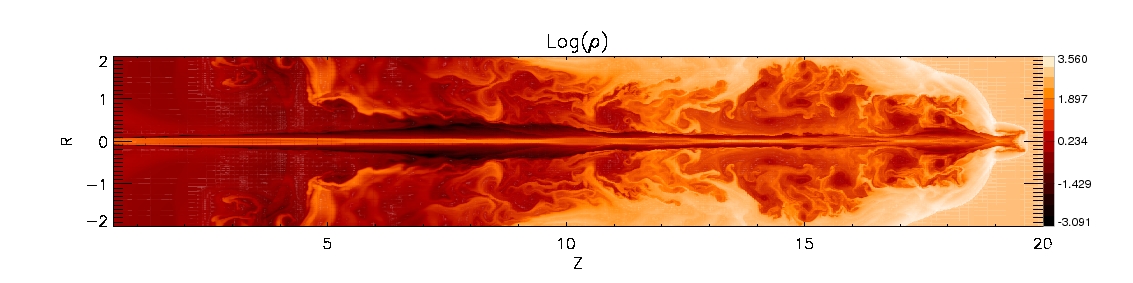}}}}
{\rotatebox{90}{\resizebox{24cm}{0.23\textwidth}{\includegraphics{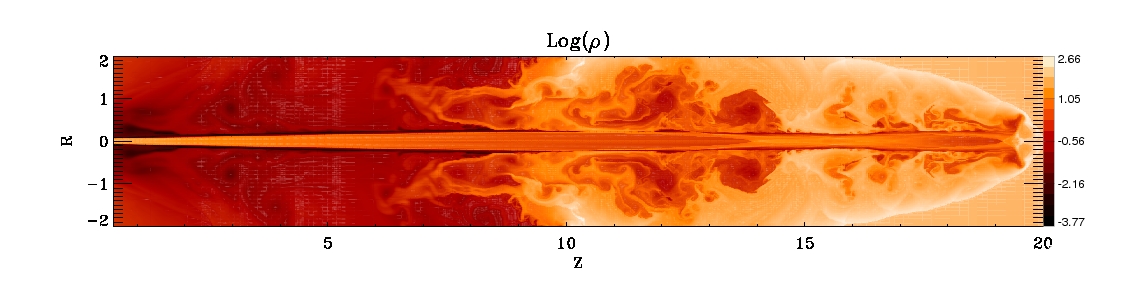}}}}
{\rotatebox{90}{\resizebox{24cm}{0.23\textwidth}{\includegraphics{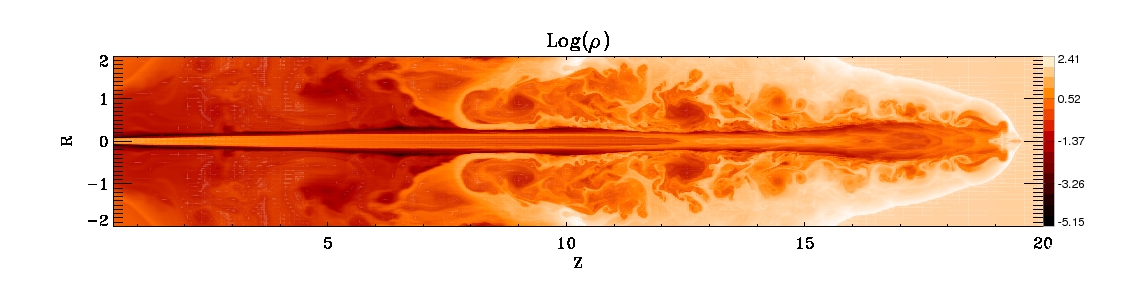}}}}
}
\caption{Contours of logarithm of density for high energy simulations: G at t=380, H at t=800, I at t= 480, J at t= 480. $R$ and $Z$ are normalised to 
($20\times R_{\rm b}$).}\label{LabelFig_Jet2DUpHighZ2}
\end{center}
\end{figure*}

In model G, the difference with model A is that the jet is slower and the upper
medium less dense. The interaction of the G jet with the external medium is therefore weaker than in A, and
the jet also remains relativistic.
At the end of simulation, this jet deposited $2.5\%$ of its energy in the 
upper medium and a fraction of the order of $1.9\%$ is reflected in the lower region.
In model H, which already had a conical expansion of the jet in the lower 
region, and when the jet reaches the interface between the two regions, the 
density of the external medium is $155.5$ higher than the density of the jet. 
When the jet goes into the upper medium, a strong bow shock forms and a high 
pressure cocoon develops. We also find a backflow that reaches a speed around 
$v_{\rm backflow}\sim - 0.5 $ at the location of the interface. 
A shock wave forms there and propagates upstream in the lower density 
region. This shock wave compresses the jet beam and collimates it 
cylindrically near the interface between the two medium.  At $t=820$, the 
this shock wave reaches the normalise distance $Z\sim 5.0$ Fig.(\ref{LabelFig_Jet2DUpHighZ2}).
In the upper medium, the high pressure cocoon induces the formation of an 
oblique shock at the jet head. This compresses and the jet radius changes from 
$R\sim 3.0\times R_{\rm jet, b}$ to $R\sim R_{\rm jet, b}/2.0$. In all, the jet decelerates abruptly from Lorentz factor $\gamma\sim 10$ to 
$\gamma\sim 2$ at this location. At this shock, the pressure becomes higher than the 
pressure in the cocoon and the state of the matter becomes relativistic and 
its effective polytropic index falls to $\Gamma\sim 1.34$. Beyond this shock the 
jet beam spreads with an angle of about $\theta\sim 6^{\circ}$ (mostly influenced by the decreasing density) and is accelerated to a Lorentz 
factor $\gamma\sim 8$ before it reaches the bow shock.
At $t=820$, the oblique shock is at normalise distance $Z\sim 11.0$ 
Fig.(\ref{LabelFig_Jet2DUpHighZ2})
At the end of simulation, the H jet deposited  $40\%$ of its energy in the 
upper lobe and a fraction of order $16\%$ is reflected in the lower region.

In the model I, the difference with the model H is that the upper medium is less dense
than the jet. In the initial periode, the jet interacts with the external medium mainly through
the front shock until the bow shock starts to develop prominently in the upper region and its pressure increases. The
high pressure of the cocoon compresses the jet and induces an internal shock in the jet that collimates 
it. Beyond this shock, the jet becomes unstable and multiple shocks develop, decelerating
the jet to Lorentz factor $\gamma\sim 8$.  At $t=480$, the oblique shock is at normalise distance $Z\sim 14.0$ Fig.(\ref{LabelFig_Jet2DUpHighZ2})
At the end of simulation, this I jet deposited $18\%$ of its energy in the 
upper lobe and a fraction of order $8\%$ is reflected in the lower region.

In model J, the difference with model H is that the jet is faster (Lorentz factor $20$ on inlet) and the external upper medium has
a density $2.5$ lower than in model H. The initial
phase of the interaction of the jet with the upper external medium 
produces also a shock wave propagating upstream, collimating the jet cylindrically.
Similar to model H, the high pressure cocoon that develops in the upper region causes a backflow
that propagates upstream. This disturbs the jet there and induces the formation
of an oblique shock that compresses, collimates, and decelerates the jet.
In fact, at the shock the jet radius now falls from 
$R\sim 4.0\times R_{\rm jet, b}$ to $R\sim R_{\rm jet, b}$. The jet is 
decelerated to a Lorentz factor $\gamma\sim 10$ through the shock. Beyond this, the jet spreads again and undergoes multiple internal 
shocks. Ultimately, this decelerates the jet beam to Lorentz factor $\gamma\sim 5$. At $t=480$, the oblique shock is at normalise distance $Z\sim 12.5$ Fig.(\ref{LabelFig_Jet2DUpHighZ2})
The faster jet J propagates for a longer distance than the jet in
H. In fact, jet J traverses a longer distance in free ballistical
propagation since the ram pressure in the jet beam
is higher. Then the oblique shock forms farther away than in H and its overall compression rate is lower. 
At the end of simulation, the jet deposited about $18\%$ of its energy in the upper lobe and a fraction of order $12\%$ gets reflected.

 \begin{figure*}
\begin{center}
\FIG{
{\rotatebox{90}{\resizebox{24cm}{0.23\textwidth}{\includegraphics{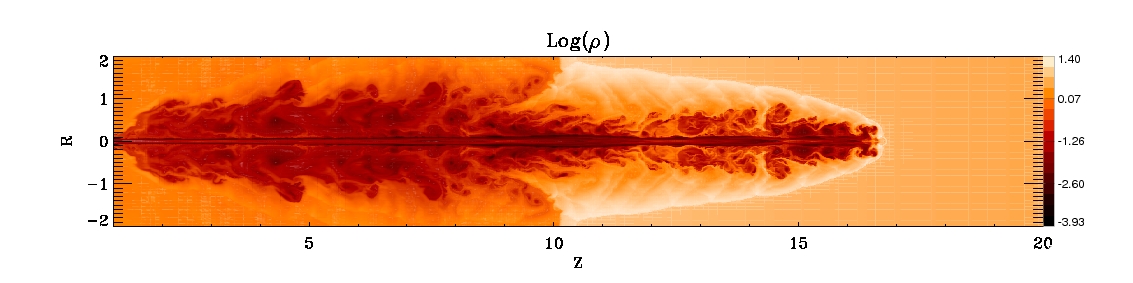}}}}
{\rotatebox{90}{\resizebox{24cm}{0.23\textwidth}{\includegraphics{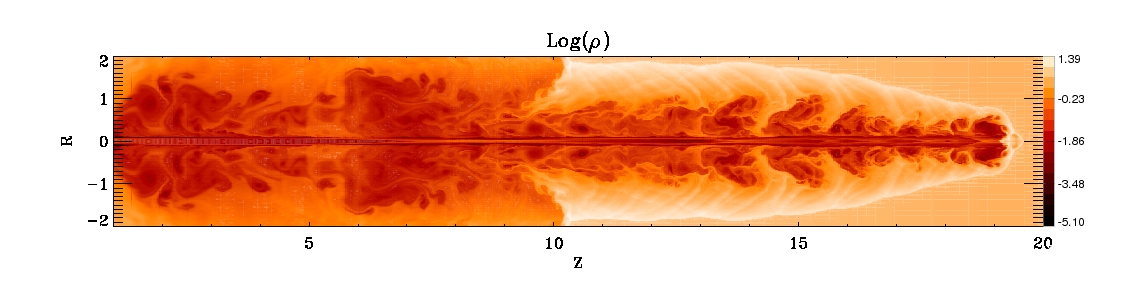}}}}
{\rotatebox{90}{\resizebox{24cm}{0.23\textwidth}{\includegraphics{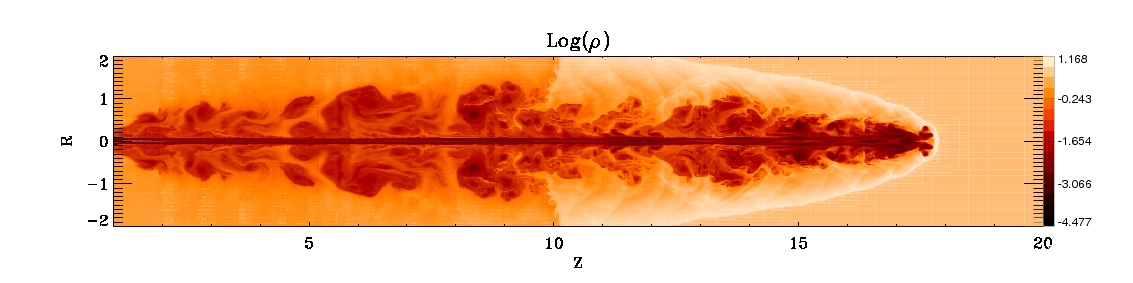}}}}
{\rotatebox{90}{\resizebox{24cm}{0.23\textwidth}{\includegraphics{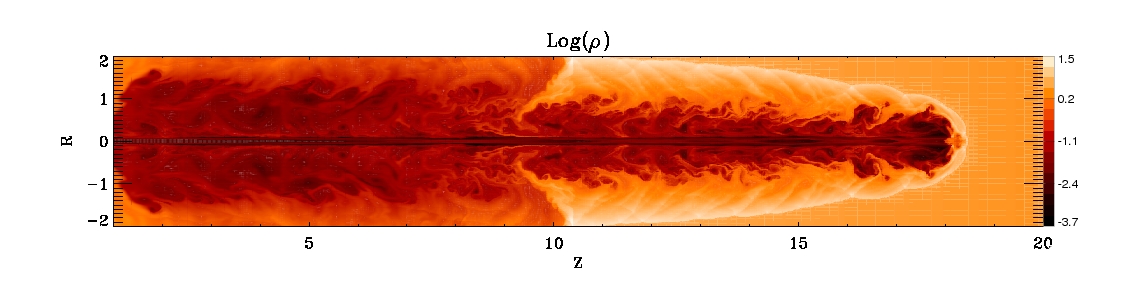}}}}
}
\caption{A contour of the logarithm of density for low energy simulations: C at t=820, D at t=820, E at t=820, F at t=820. $R$ and $Z$ are normalised to 
($20\times R_{\rm b}$).}\label{LabelFig_Jet2DUpHighZ2B}
\end{center}
\end{figure*}

\subsubsection{Summary of all cases}

\begin{figure}
\begin{center}
\FIG{
{\rotatebox{0}{\resizebox{0.9\columnwidth}{5cm}{\includegraphics{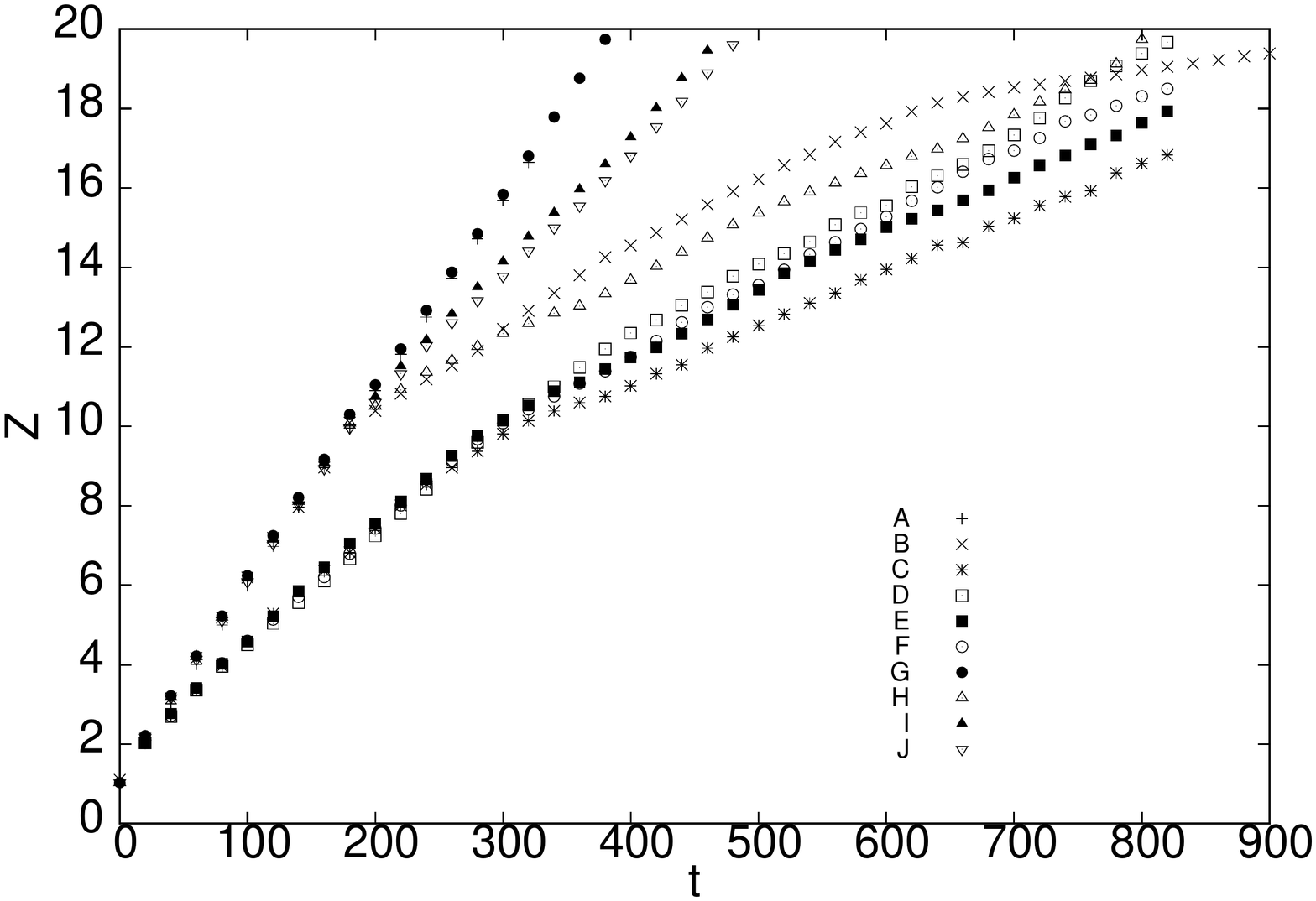}}}}
}
\caption{The position of the jet head as function of time, for all cases.}\label{JetHeadCompart}
\end{center}
\end{figure}

Figures~\ref{LabelFig_Jet2DUpHighZ2}-\ref{LabelFig_Jet2DUpHighZ2B} gives an impression of the jet morphologies, at the end of the simulations.
In our models A, G with high energy and upper medium of about the jet density, we end up with a relativistic jet in both regions (inner/upper).
In the upper, slightly overdense region (for A), the interaction with the external medium is stronger than
in the underdense inner region, and we find the formation of a relatively smooth cocoon. 
In the model B with high energy and a much higher density in the upper medium, the jet is relativistic in the lower region, where it is seen to 
propagate with a Lorentz factor of about $5$, and turns to a sub-relativistic jet 
with propagation speed $0.3$ in the very dense upper region. 
In all models with low energy C, E, the jet remains relativistic in the lower region despite fairly turbulent cocoons, and
is strongly decelerated when it crosses the interface between the two media. In the cases undergoing strong deceleration, a shock wave propagates upstream 
(a reflection from impacting the higher density environment), and this compresses and decelerates the jet.
In the model with an initial opening angle, an oblique shock forms before the jet
reaches the contact interface between the two media for the cases D, F with low energy.
This shock survives the interaction with, and gets enhanced by the shock wave propagating upstream that forms when the jet interacts with the upper
medium. In conical models H, I, J with higher energy, such an oblique shock forms only when the jet starts to interact with the higher density upper medium.  
These pronounced differences in the propagation behavior between the two jets, is clearly seen
when we quantify the propagation of the jet head, which is collected as function of time in 
Fig.~\ref{JetHeadCompart}, for all models. We can conclude that this is relevant to the FR I/FR II dichotomy: we find FR II behavior where the energy of the 
jet is high and the density of the external medium is not high enough to significantly impart a strong reflected shock, and FR I behavior when it can,
even if the density beyond the interface decreases and also if the jet is not collimated. 

Finally, we mention that in all these simulation, when the jet crosses the intersection between
the two medium, we witness the appearance of Richtmyer-Meshkov instabilities at the shocked contact interface.
We were able to resolve the local vortical development, using high resolution. In fact, the 
Richtmyer-Meshkov instabilities evolve out of vorticity deposition on the interface, which is due to the propagating jet bow shock
passing it.

\section{Conclusions}
In this paper, we extended, presented and applied the relativistic 
hydrodynamics AMRVAC code \citep{Melianietal07,vanderHolst07} with a 
generalized variable polytropic index equation 
of state for the purpose of modeling the relativistic shocks in GRBs and AGN 
jets. As follows in the appendix, we used various shock-capturing schemes on a variety of test cases for
adiabatic and non-adiabatic cases for code validation. 
We demonstrated the code ability with stringent new test problems, developed 
according to the astrophysical context. We tested the code also for a case with
a heated flow, using the Synge-like equation of state from \cite{Melianietal04}.
The Riemann problems were also solved exactly, and the AMR simulations handled 
cases with Lorentz factors of order 100 accurately.

We explored a new scenario for sudden deceleration of relativistic
jets in the FR I radio galaxies. 
This model for jet deceleration is based on a density
jump in the external medium, with density suddenly increasing to the upper medium. We include
models of conical jets with an initial opening angle, and allow for decreasing denisty profiles within the upper medium.
We investigated the propagation and 
the dynamics of relativistic AGN jets over a long distance, 
resolving small scale instabilities which develop in the cocoon and 
which could be responsible for particle acceleration. This study was  
only possible with adaptive mesh refinement.
We quantified and discussed the deceleration of the jet resulting from its 
interaction with a layered medium. We point out that jet deceleration can be significantly aided
by an internal oblique shock that forms in jets with conical expansion.
Under fairly extreme (but not as extreme as pursued in other simulations to date) density conditions in the dense upper region, it can reproduce the strong deceleration observed in FR I jets. 

In the early phase, the head of the cold fast (beam Lorentz factor 10 or 20) jet propagates in the low density medium at a 
high Lorentz factor $\gamma=5$. 
In cases with high energy, almost no cocoon or backflows develop in this phase, and as the jet is heavier than the external 
region, it behaves more ballistically. When observed in this region, the relativistic beaming is high, so that one would observe a one-sided jet.
In low energy jets, a cocoon always forms and slow backflows develop. They in turn induce multiple
internal shocks making the jet decelerate and re-accelerate. In jets with low energy and with an initial conical flow, the
flow remains ballistic until it reaches a self-consistently forming oblique shock.
This compresses and decelerates the jet and two regions appear in the jet beam, the first with high relativistic beaming, and the second with low Lorentz 
factor, showing clear disturbances by the surrounding cocoon.  

In the outer region where the density increases suddenly, we find increased 
entrainment of ambient material through velocity shear instabilities at the jet boundary and working surfaces. 
Note that we here for the first time address how 
the prior interaction of the jet with the low density ambient medium in the inner region (which makes the head of the jet consist of swept-up ambient medium and a shocked beam) affects jet propagation in denser media. 
The pre-structured jet head has lower Mach number, which gives rise to a strong interaction 
with the denser ambient medium. 

For models with a weak increase in density in the external
medium, the cocoon shows little turbulent structure and no backflow appears. The jet in this case remains
stable and relativistic in the denser upper region. Also a fraction of its energy is deposited in the 
external medium, but only through the frontal shock. This model could be relevant for jets in FR II.
With this scenario, one can explain how most of the energy of the jet is 
deposited in the outer region, since relatively little jet energy gets 
transferred to the medium in the inner region. 

Those models with more extreme increases in density in the external
medium, showed the formation of an overpressured and turbulent cocoon. A strong
backflow develops in the outer region, which disturbs the jet
structure. The backflow compresses the jet and induces internal shocks which
give rise to knot formation. The jet  becomes subrelativistic within the simulated domain. 
The models with initial conical jets show the development of an oblique shock
that decelerates the jet.
In future work, we will follow these jets over increasingly larger distances in 3D scenarios. We also intend to provide synthetic
observational radio maps from our computed jet models.

\appendix\label{appendixA}
\section{Relativistic hydrodynamics and the equation of state}

The special relativistic hydrodynamic evolution of a perfect fluid is governed
by the conservation of the number of particles, and energy-momentum conservation. These two conservation laws can be written as
\begin{eqnarray}
\left(\rho\,u^{\mu}\right)_{\mu}=0\,,\;\;\;
\left(T^{\mu\nu}\right)_{\mu}=0\,,
\end{eqnarray}
where $\rho$, $\vec{u}=\left(\gamma,\gamma\,\vec{v}\right)$, and
$T^{\mu\nu}=\rho\,h\,u^{\mu}\,u^{\nu}+\,p\,g^{\mu\nu}$ define, respectively, the proper
density, the four-velocity and the stress-energy tensor of the perfect fluid. Proper density is related to
the number density $n$ in the fluid rest frame $\rho=n m_p$, where $m_p$ indicates the particle (proton) rest mass. 
The definitions involve
the Lorentz factor $\gamma$, the fluid pressure $p$, and the relativistic specific enthalpy
$h$. For the (inverse) metric
$g^{\mu\nu}$, we take the Minkowski metric. Units are taken where the light speed equals unity.

These equations can be written in conservative form involving the Cartesian coordinate axes and the time axis of a fixed
`lab' Lorentzian reference frame as
\begin{eqnarray}
\frac{\partial U}{\partial t} +
\sum^{3}_{j=1}\frac{\partial F^{j}}{\partial x^{j}}=0\,.
\end{eqnarray}
The conserved variables can be taken as
\begin{equation}
U=\left[D=\gamma\,\rho,\vec{S}=\gamma^2\rho\,h\,\vec{v},
\tau=\gamma^2\rho\,h-p-\gamma\rho\right]^{T}\,,
\end{equation}
and the fluxes are then given by
\begin{equation}
F=\left[\rho\gamma\,\vec{v},\gamma^2\rho\,h\,\vec{v} \vec{v}+p\,{\bf I},
    \gamma^2\rho\,h\vec{v}- \gamma \rho\, \vec{v}\right]^{T}\,,
\end{equation}
where ${\bf I}$ is the $3\times 3$ identity matrix.
To close this system of equations, we can use the Synge-like equation of state (EOS) for an ideal
gas as also used by \cite{Melianietal04}, which is a polytropic equation with a 
corresponding classical polytropic index $\Gamma$ relating
\begin{equation}
p\,=\,\left(\frac{\Gamma - 1}{2}\right)\,\rho \,\left(\frac{e}{m_p}-\frac{m_p}{e}\right)\,,
\end{equation}
where $e=m_p+e_{\rm th}$ is the specific internal energy including rest mass, and $e_{\rm th}$ 
is the specific thermal energy.
This specific internal energy is given in function of the pressure and density by rewriting the above equation to
\begin{equation}
\frac{e}{m_p}=\frac{1}{\Gamma - 1}\frac{p}{\rho}+
\sqrt{\left(\frac{1}{\Gamma - 1}\frac{p}{\rho}\right)^2+1}\,.
\label{inte}
\end{equation}
The relativistic specific enthalpy is then given by
\begin{equation}
h=\frac{1}{2}\,\left(\left(\Gamma +1\right) \frac{e}{m_p} -\left(\Gamma -1\right)\frac{m_p}{e}\right)\,.
\label{enth}
\end{equation}
The sound speed (in light speed units) is then found from
\begin{eqnarray}
c_{\rm s}^2&=&\frac{1}{h}\frac{{\rm d}p}{{\rm d}\rho}\nonumber\\
&=&\frac{1}{h}\,\frac{p}{\rho}
\left(\frac{\left(\Gamma+1\right)}{2}+\frac{\left(\Gamma-1\right)}{2}\frac{m_p^2}{e^2}\right)\,.
\end{eqnarray}
At each time step in the numerical integration, the primitive variables $(\rho,\vec{v},p)$ involved in flux expressions
should
be derived from the conservative variables $U$ resulting in a system of
nonlinear equations.  One can bring this system into a single equation for the pressure
$p$, directly following from the definition of the conserved variable $\tau$ from
\begin{equation}\label{NR}
(\tau+p+D)\,(1-v(p)^2)\,-\,\rho h(p)=0\,,
\end{equation}
which, once solved for $p$ yields $\vec{v}={\vec{S}}/{(\tau+p+D)}$. One inserts $h$ from Eq.~(\ref{enth}), in which
one uses Eq.~(\ref{inte}), with in addition $D=\gamma \rho$ while $1/\gamma^2=1-S^2/(\tau+p+D)^2$.
This nonlinear equation~(\ref{NR}) is solved using a
Newton-Raphson algorithm.
For a chosen fixed index $\Gamma=5/3$, we then achieve a locally varying, effective polytropic
index taking on values between its relativistic $4/3$ and classical $5/3$ extremes, found from
\begin{equation}
\Gamma_{\rm eff} = \Gamma -\frac{\Gamma-1}{2}\left(1-\frac{m_p^2}{e^2}\right)\,.
\label{eff}
\end{equation}
This provides an excellent approximation to the true Synge gas expression, while avoiding costly Bessel function
evaluations. If $1<\Gamma\le 5/3$, we incorporate effects ranging from near-isothermal conditions to adiabatic flow, and if $\Gamma>5/3$ we incorporate the effect of energy losses.

\section{Numerical tests for realistic EOS}

The main aim of this appendix is to point out that we have determined the exact solutions to newly selected Riemann problems with a realistic equation state.
 This exact solution is here used, and can be used by other authors, to benchmark codes. 
We test the code for adiabatic EOS Riemann problems, as also investigated by \citet{Mignone&McKinney07}, and for a more general case with equivalent 
classical polytropic index $3/2$ to mimic the heating in a relativistic fluid \citep{Melianietal04}.
 In each test, we analyse the effect of the equation of state with variable effective polytropic index on the behaviour of the shock and rarefaction waves.  
We include tests with high Lorentz factor $100$, with the aim to investigate shocks in Gamma Ray Bursts, where the forward shock is always relativistic and 
the reverse shock Newtonian. The matter shocked by the forward shock has then an effective polytropic index reaching $4/3$. However, the matter shocked by 
the reverse shock has effective polytropic index $5/3$. This difference state of the matter between the two shocked media induces a change in the propagation speed of the shocks. In fact, using this EOS will improve the simulations for GRBs and induce some variation from the results in \cite{Melianietal07}.

\begin{figure*}
\begin{center}
\FIG{
{\rotatebox{0}{\resizebox{0.24\textwidth}{5cm}{\includegraphics{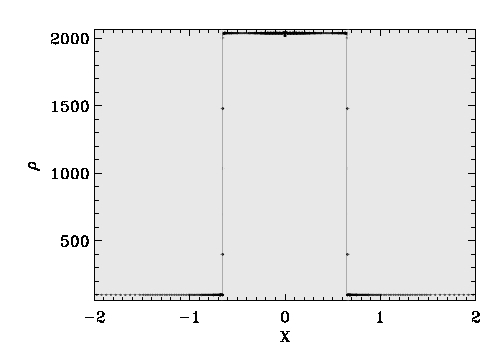}}}}
{\rotatebox{0}{\resizebox{0.24\textwidth}{5cm}{\includegraphics{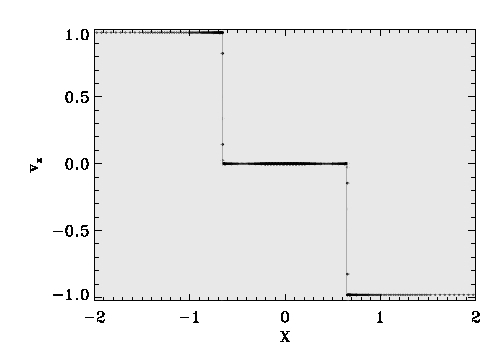}}}}
{\rotatebox{0}{\resizebox{0.24\textwidth}{5cm}{\includegraphics{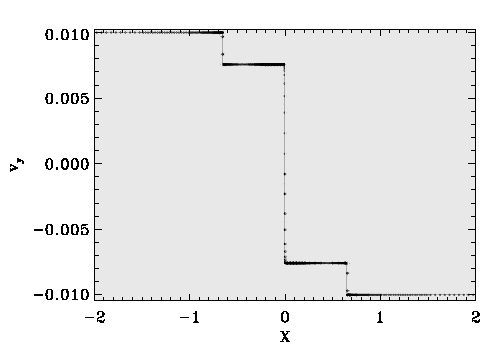}}}}
{\rotatebox{0}{\resizebox{0.24\textwidth}{5cm}{\includegraphics{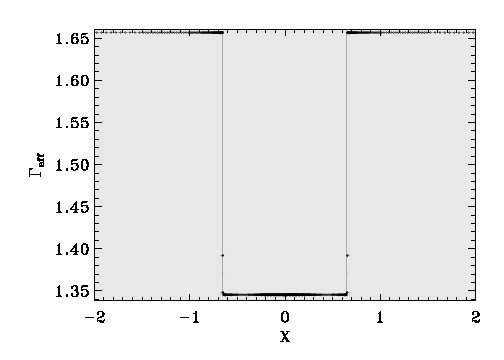}}}}
}
\caption{One-dimensional Riemann problem for colliding, identical cold flows, with a jump in tangential velocity $v_y$. The solid lines are the exact solution. The result is plotted at $t=2$.}\label{LabelFig_Rie1DA}
\end{center}
\end{figure*}

\begin{figure*}
\begin{center}
\FIG{
{\rotatebox{0}{\resizebox{0.24\textwidth}{5cm}{\includegraphics{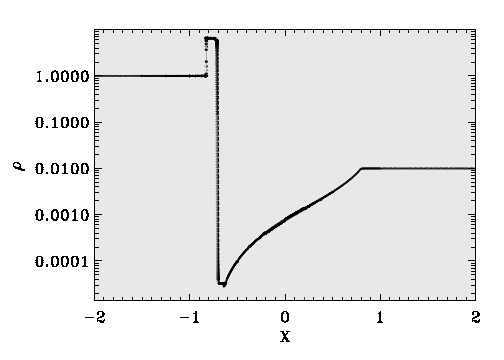}}}}
{\rotatebox{0}{\resizebox{0.24\textwidth}{5cm}{\includegraphics{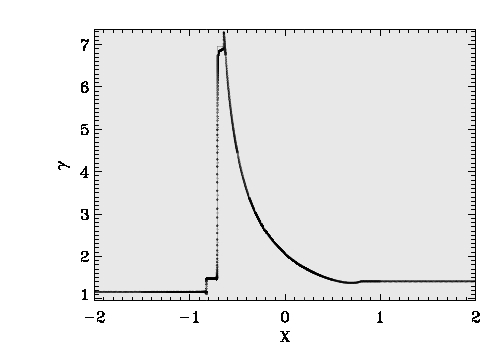}}}}
{\rotatebox{0}{\resizebox{0.24\textwidth}{5cm}{\includegraphics{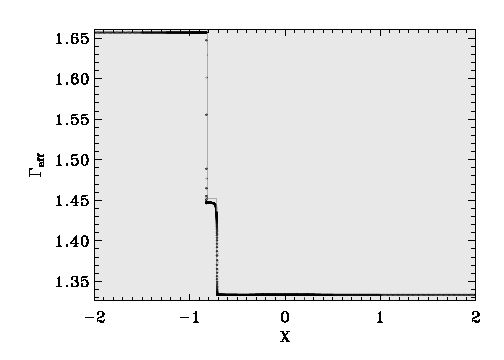}}}}
{\rotatebox{0}{\resizebox{0.24\textwidth}{5cm}{\includegraphics{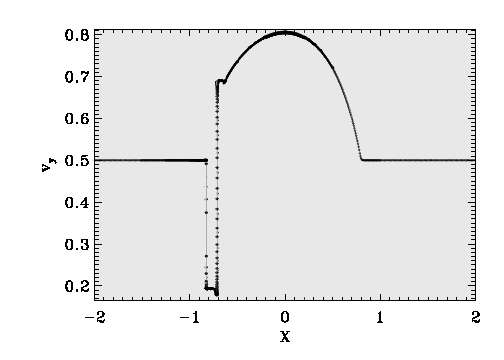}}}}
}
\caption{One-dimensional Riemann problem for two states with different initial 
thermodynamic state. The fluid at left has a classical state and the fluid 
at right a relativistic state. The solid lines are the exact 
solution. The result is plotted at $t=1$.}\label{LabelFig_Rie1DB}
\end{center}
\end{figure*}

\subsection{Exact solution to Riemann problems with realistic EOS}\label{ExactRiemann}

The solution of the one-dimensional Riemann problem in hydrodynamics
consists of determining the temporal evolution of a fluid which, at
some initial time, has two adjacent uniform states characterized by
different values of uniform velocity, pressure and density. These
initial conditions completely determine the way in which the
discontinuity will decay after removal of the barrier separating the
initial ``left'' and ``right'' states.

In general, the Riemann problem requires the solution of a nonlinear
algebraic system of equations written as a function of a set of
unknown quantities. In relativistic hydrodynamics (RHD) an exact
solution has been obtained only rather recently and was proposed
by~\cite{marti94} for flows that are purely along the direction normal
to the initial discontinuity. This work has then been extended to the
case in which tangential velocities are present (\cite{Ponsetal00}) and
improved in efficiency by exploiting the relativistic invariant
relative velocity between the two states to predict the wave pattern
produced (\cite{rezzolla01} and~\cite{rezzolla03}).

The exact solution is found after expressing all of the quantities
behind each wave as functions of the value of the unknown gas pressure
$p$ at the contact discontinuity. In this way, the problem is reduced
to the search for the value of the pressure that satisfies the jump
conditions at the contact discontinuity.
Recently~\cite{Giacomazzo&Rezzolla06} found for the first time (after a
first attempt by~\cite{romeroetal05} limited to a very special case) a
general procedure to compute the exact solution of the Riemann problem
in relativistic magnetohydrodynamics (RMHD). This solver can also
be used in RHD by setting to zero the magnetic field and in this case
it is similar to the method developed by~\cite{Ponsetal00}.
Moreover, the exact solver can implement,
both in RHD and in RMHD, various equations of state of the form
$p=p(\rho,e_{\rm th})$ making it a very general tool that has been used for the
testing of several special and general relativistic numerical
codes. To the best of our knowledge, 
this is the first time in which exact solutions
with generic initial states are computed with a different than polytropic EOS.

\subsection{Numerical schemes}

We include in this section various stringent tests to validate the code. We 
perform one-dimensional test problems to show the code ability to resolve Riemann 
problems, and show that we reproduce the exact wave patterns in special 
relativistic hydro with varying EOS.
We use different discretization schemes, namely TVDLF (T\'oth \& Odstr\v cil 1996), 
HLLE \citep{Hartenetal83, Einfeldt88}, HLLC (Mignone \& Bodo 2005), and also a hybrid mixture between TVDLF and 
HLLC. The latter ensures that in regions where spurious oscillations can be 
induced by HLLC, the code switches to TVDLF. This switch is effective 
in cells where the fluxes as computed at left and right edge change direction. 
For the linear reconstruction from cell center to edge, we use a robust minmod scheme that reduces any spurious numerical oscillations. All these 1D tests are compared to their exact solution, computed as outlined
in Section~\ref{ExactRiemann}.

\subsection{1D Tests: Riemann problems}

\begin{figure}
\begin{center}
\FIG{
{\rotatebox{0}{\resizebox{\columnwidth}{5cm}{\includegraphics{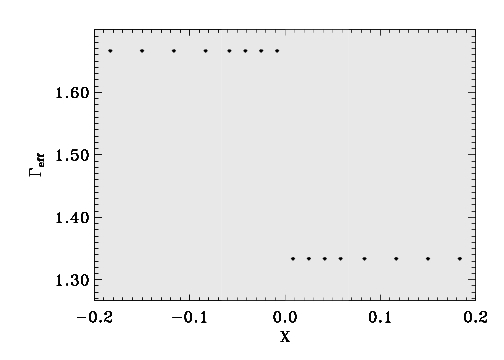}}}}
}
\caption{Riemann problem for two states with different initial 
thermodynamic state, with a stationary contact discontinuity at $x=0$.}\label{LabelFig_Rie3DB}
\end{center}
\end{figure}

\begin{figure}
\begin{center}
\FIG{
{\rotatebox{0}{\resizebox{0.49\columnwidth}{4cm}{\includegraphics{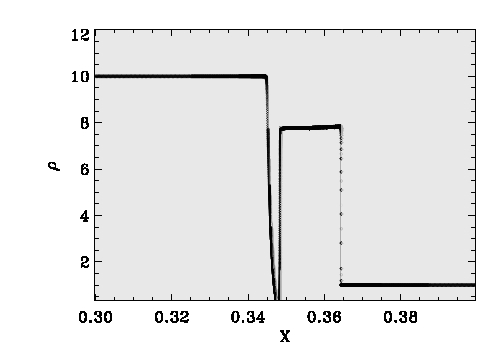}}}}
{\rotatebox{0}{\resizebox{0.49\columnwidth}{4cm}{\includegraphics{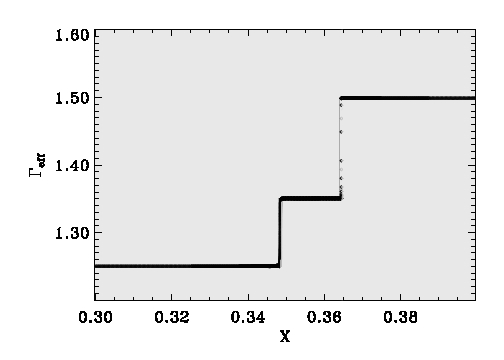}}}}
}
\caption{One-dimensional Riemann problem for the internal shock in a heated wind,
colliding with slower outflow, plotted at $t=0.4$.}
\label{LabelFig_Test5}
\end{center}
\end{figure}

We start with the collision of two cold flows with, in addition, an opposite orientation of
tangential velocity. The initial states are $\rho_{\rm L} =100, \, p_{\rm L} = 1,\,v_{\rm x,\,L}=5/\sqrt{26}, v_{\rm y,\,L}=0.01$ (left) and  $\rho_{\rm R} =100, \, p_{\rm R} = 1,\,v_{\rm x,\,R}=-5/\sqrt{26},\, v_{\rm y,\,R}=-0.01$ (right). Both flows have a Lorentz factor 
$\gamma\sim 5$ and the classical polytropic index $5/3$.  The 
simulation is done with $20$ cells at the lowest grid level, and we allow for 
$7$ levels on the spatial range $-2<x<2$. The test shows the characteristic pattern of two 
shock waves and a stationary discontinuity in the tangential velocity. The two 
shocks convert the kinetic energy to thermal energy and the effective polytropic 
index which was $\Gamma_{\rm eff}=1.657$ throughout drops locally to $1.345$ increasing 
the compression rate $\rho_{\rm C}/\rho_{\rm L} \sim 20.44$, where 
$\rho_{\rm C}$ denotes the central proper density. 
In Fig.~\ref{LabelFig_Rie1DA}, we plot the result at $t=2$ as obtained with the 
hybrid version of HLLC. This scheme captures the two shock waves accurately, as 
they are propagating with a shock speed $v_{\rm sh} \sim 0.33$, and seperate the two 
regions with different states of the matter. Indeed, a classical fluid exists in 
front of the shock waves and an ultra-relativistic state is found between the 
shock waves.

In the second test, we follow the evolution of an initial discontinuity
between two fluids with different thermodynamic states, both moving to the right. 
At right, the relativistic fast gas is characterized by 
$p_{\rm R}=5000,\,\rho_{\rm R}=0.01,\, v_{\rm x, R}=0.5,\,v_{\rm y, R}=0.5$, and this interacts with
the cold slower gas at left 
$p_{\rm L}=0.01,\,\rho_{\rm L}=1,\, v_{\rm x, L}=0.1,\,v_{\rm y, L}=0.5$, thereby creating a flow
to the left. We take $\Gamma=5/3$, and obtain at the left an effective polytropic index $\Gamma_{\rm eff, L}=5/3$ and at right, it is 
$\Gamma_{\rm eff, R}=4/3$. The simulation is done with $20$ cells at the lowest grid level, and we 
allow for $9$ levels on the spatial range $-2<x<2$. The evolution leads to the formation of a
shock wave propagating to the left, which heats and compresses the fluid. 
Between this shock and the contact discontinuity, the effective polytropic index 
of the compressed fluid drops to $1.45$. In Fig.~\ref{LabelFig_Rie1DB}, we plot 
the result at $t=1$ as obtained with the hybrid version of HLLC.
The compression rate $\rho_{\rm L, s}/\rho_{\rm L} \sim 6.69$, where $\rho_{\rm L, s}$ denotes the density of the compressed fluid.
More to the right of the contact, a rarefaction wave propagates into the 
relativistic fast fluid. A numerical difficulty is produced by the strong drop 
in the mass flux, even though the tangential velocity increases at the contact 
discontinuity, such that the Lorentz factor between
the contact discontinuity and the tail of the rarefaction wave is $4.63$ higher than in front of the contact discontinuity (see Fig.~\ref{LabelFig_Rie1DB}).
In this test, it is vital to introduce a realistic equation of state 
with effective polytropic index becoming function of the temperature. As the shock
propagates to the left, it produces a mildly-relativistic adiabatic compression of the
flow, where the state of the fluid is described by $4/3<\Gamma<  5/3$.
This state can not be analysed with a classical constant polytropic equation of state.

The third test represents an isolated contact discontinuity with only a jump
in the density $\rho$ such that there is also a jump in the effective polytropic 
index. The initial states are $\rho_{\rm L} = 10^4$, $p_{\rm L}=1.0$, 
$v_{\rm x, L}=0.0$, $v_{\rm y, L}=0.4$ (left) and $\rho_{\rm R} = 0.125$, $p_{\rm R}=1.0$, 
$v_{\rm x, R}=0.0$, $v_{\rm y, R}=0.4$ (right). The simulation is done with
$12$ cells at the lowest grid level, and we allow for 2 levels on spatial grid 
$-0.2<x<0.2$. In Fig.~\ref{LabelFig_Rie3DB}, we plot the result (at $t=2$, representative of all times) as obtained with the HLLC 
scheme. The test shows that HLLC resolves an isolated stationary 
discontinuity exactly (and thus the isolated jump in effective polytropic index).

In the fourth test, we follow the evolution of an initial discontinuity between
two flows to the right, with different Lorentz factor, both having a 
polytropic index $\Gamma=3/2$. This represents the critical value in the Parker model for the solar wind \citep{Parker60}, and due to our EOS, both flows undergo a heating. 
The initial states are $\rho_{\rm L}=10.0$, $p_{\rm L}=100.0$, $v_{\rm x, L}=0.866025$, 
$v_{\rm y, L} = C_{\rm s,L}$ (left) and  $\rho_{\rm R}=1.0$, 
$p_{\rm R}=10^{-3}$, $v_{\rm x, R}=0.30491$, $v_{\rm y, R} = C_{\rm s,R}$ 
(right) where $C_{\rm s}$ denotes the sound speed. 
The Lorentz factor at left is already order 100 here.
The aim of this test is to mimick the internal 
shock between a mildly relativistic outflow and subsequent hot and fast ejecta. 
The simulation is 
done with $600$ cells at the lowest grid level, and we allow for $13$ levels 
on spatial grid  $-2.0<x<2.0$ with a discontinuity at $x=0.0$.
In Fig.\ref{LabelFig_Test5}, we plot the result at $t=0.4$. In this test, 
we succeed to resolve the forward shock propagating to the right, the contact 
discontinuity, and the rarefaction wave propagating to the left (as the fluid
in the left is hot). The left propagating rarefaction wave increases the
normal speed to $v_{\rm x, L}\sim 0.879$, and together with the transverse speed the Lorentz factor reaches 
the value $\gamma\sim 233$. 
This high Lorentz factor in a thin shell between the contact discontinuity and
the tail of the rarefaction wave represents a very difficult numerical problem 
and requires the use of AMR.

\begin{figure*}
\begin{center}
\FIG{
{\rotatebox{0}{\resizebox{0.3\textwidth}{5cm}{\includegraphics{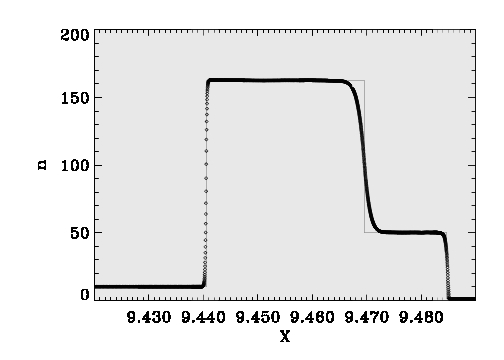}}}}
{\rotatebox{0}{\resizebox{0.3\textwidth}{5cm}{\includegraphics{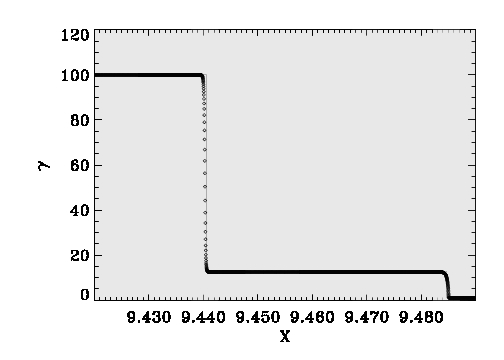}}}}
{\rotatebox{0}{\resizebox{0.3\textwidth}{5cm}{\includegraphics{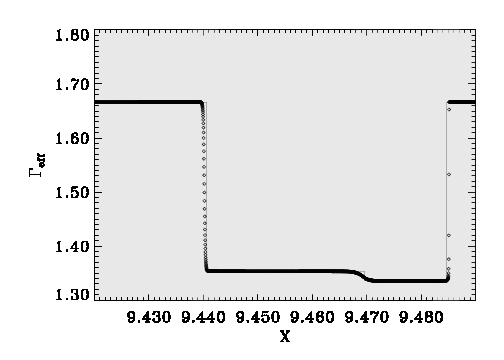}}}}
}
\caption{GRB test case: a cold Lorentz factor 100 flow penetrates into cold static medium. The solid lines are the exact solution. Left: the density, center: Lorentz factor, right: effective polytropic index.}
\label{LabelFig_GRBTest100}
\end{center}
\end{figure*}

\begin{figure*}
\begin{center}
\FIG{
{\rotatebox{0}{\resizebox{0.3\textwidth}{5cm}{\includegraphics{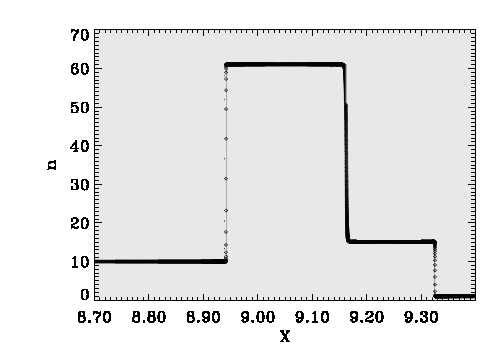}}}}
{\rotatebox{0}{\resizebox{0.3\textwidth}{5cm}{\includegraphics{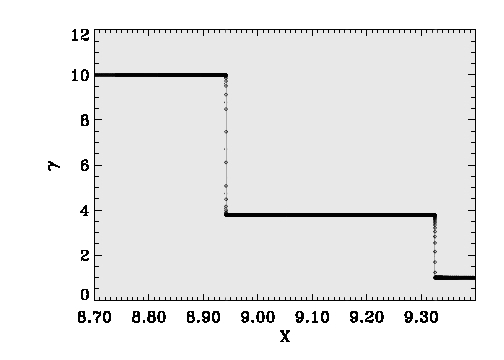}}}}
{\rotatebox{0}{\resizebox{0.3\textwidth}{5cm}{\includegraphics{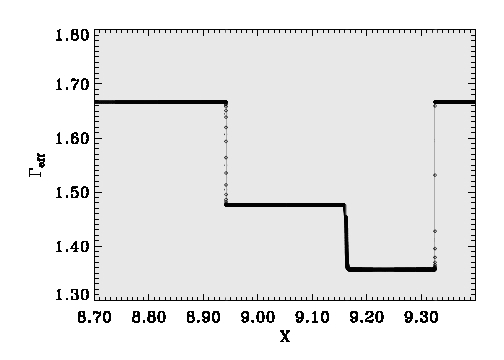}}}}
}
\caption{Second test case: a cold Lorentz factor 10 flow penetrates into cold static medium. The solid lines are the exact solution. Left: the density, center: Lorentz factor, right: effective polytropic index.}
\label{LabelFig_GRBTest10}
\end{center}
\end{figure*}

\subsection{1D Riemann problem tests for Gamma Ray Bursts}\label{grb}

We now test the ability of the code to model the ultra-relativistic shocks 
between a relativistic cold flow with a cold external medium. This interaction 
induces a strong compression of the swept up matter from the external medium. 
We model three such cases, two with a Lorentz factor of 100 and a third with a 
Lorentz factor of 10. In all cases, we will find a reverse shock, contact 
discontinuity, and forward shock configuration. 
In these tests the density of the fast flow
is 10 times higher than the density of the stationary medium. All the parameters of the tests are
given in Table~\ref{table:2}.
These extreme conditions are representative of GRB dynamics, and any code 
tailored for studying ultra-relativistic flows must be able to compute these 
situations accurately. All these test are done  with the hybrid version of 
HLLC. In all these tests there are all four regions
that characterize the interaction between an outward moving
relativistic beam and the cold external medium: (1) the external medium at rest, 
(2) the shocked external medium, (3) part of the beam material which is shocked by the reverse
shock, (4) unshocked cold material of the beam.

\begin{table}
\caption{GRB test parameters, the Lorentz factor and the transverse speed of 
left flow, and the density ratio between left and right state. The  fluid in the right is initially at rest.}              
\label{table:2}      
\centering                                      
\begin{tabular}{c c c c}          
\hline\hline                        
Test & $\gamma_{\rm L}$ & $\rho_{\rm L}/\rho_{\rm R}$ & $v_{\rm y, L}$ \\    
\hline                                   
    1 & 100 & 10 & 0.0 \\      
    2 & 10 & 10    & 0.0 \\
    3 & 100 & 10  & 0.01 \\
\hline                                             
\end{tabular}
\end{table}

In the first test, the simulation is done with $256$ cells at the lowest grid level, and we 
allow for $12$ levels on the spatial range $-0.1<x<10$ with discontinuity at 
$x=0.0$, the result at $t=9.5$ is shown in Fig.~\ref{LabelFig_GRBTest100}. The compression rate of the density by the forward shock is 
$\gamma_{2}\,n_{2}/n_{1}=\gamma_{2}\left(4\,\gamma_2+3\right)\sim 669.73$ 
and the thermal energy reaches 
$e_{2}/m_{p} = \left(\gamma_{2}-1\right)n_{2}\sim 616.86\,n_{1}$
as the Lorentz factor after the forward shock is 
$\gamma_{2} = \frac{\gamma_{4}^{1/2} f^{1/4}}{\sqrt{2}}\sim 12.57$ where $f=\frac{n_{4}}{n_{1}}=10$ \citep{Sari&Piran95}.
In the region of the shocked swept-up matter, the polytropic index is $4/3$ as the thermal
energy is higher than the mass energy. Moreover, the reverse shock is propagating 
with a relative Lorentz factor 
$\bar{\gamma_{3}}\sim\sqrt{\frac{\gamma_{4}}{2\,f^{1/2}}}\sim 3.97$, hence the matter
is compressed according to
$n_{3}/n_{4}\sim \left( 4\,\bar{\gamma_3}+3\right) \sim 18.9$ (the exact obtained value is in fact
somewhat lower)
and the thermal energy enhanced 
$e_{3}/m_{p}=\left(\bar{\gamma_{3}}-1\right)\,n_{3}=56.07\,n_{4}$. This implies that the state of the matter is also
relativistic and the polytropic index $\Gamma_{3}\sim 4/3$. In this case both the forward
shock and the reverse shock are relativistic. The numerical difficulty is to resolve
the very thin region resulting from the strong compression of swept-up matter, which
has an extent of 
$\Delta X_{2}(t) = 0.0015  t$. It is easier to resolve
the region between the reverse shock and the contact discontinuity, as the compression 
by the reverse shock is weaker, and this region extends over 
$\Delta X_{3}(t) = 0.00315  t$.
\begin{figure*}
\begin{center}
\FIG{
{\rotatebox{0}{\resizebox{0.3\textwidth}{5cm}{\includegraphics{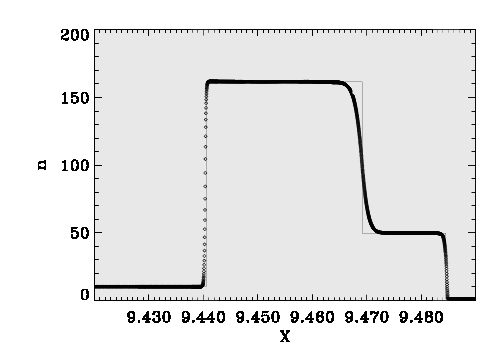}}}}
{\rotatebox{0}{\resizebox{0.3\textwidth}{5cm}{\includegraphics{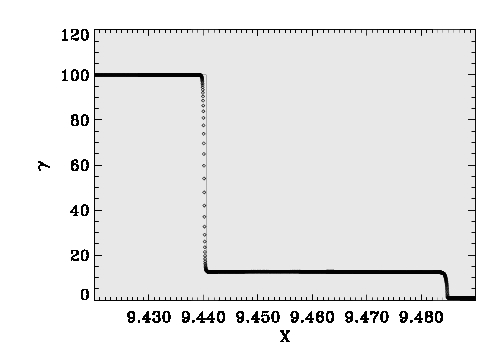}}}}
{\rotatebox{0}{\resizebox{0.3\textwidth}{5cm}{\includegraphics{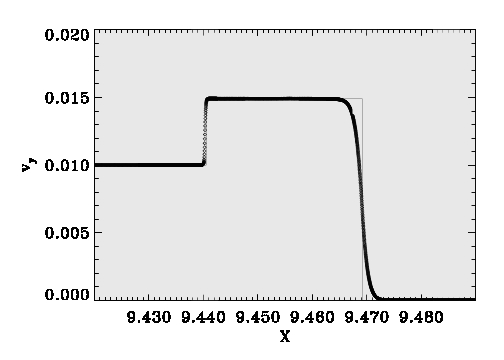}}}}
}
\caption{GRB test case with transverse velocity: a cold Lorentz factor 100 beam with also a small transverse velocity flow, penetrates into cold static medium. The solid lines are the exact solution. Left: the density, center: Lorentz factor, right: transverse velocity.}
\label{LabelFig_GRBTest100Tr}
\end{center}
\end{figure*}

In a second test, a relativistic cold flow with a Lorentz factor $10$ interacts with a cold external medium.
This case corresponds more to the interaction of an
AGN or micro-quasar jet with external medium than with a GRB case.
The simulation is done with $256$ cells at the lowest grid level, and we 
allow for $8$ levels, the result at $t=9.5$ is shown in Fig.~\ref{LabelFig_GRBTest10}.
With the weak efficiency of the forward shock $\gamma_{2}\sim 3.79$ to compress, 
the thermal energy of swept-up matter increases less and the effective polytropic
index reaches a value $\Gamma_{\rm eff}=1.356$. The compression rate in this 
case is lower and hence the swept-up matter has an extent
of $\Delta X_{2}(t) = 0.0163 t$, which allows to resolve easily the contact discontinuity
and the forward shock. The reverse shock is mildly relativistic with
a relative Lorentz factor $\bar{\gamma_{3}}\sim 1.53$  such
that the effective polytropic index of the shocked beam drops only to $\Gamma_{\rm eff}=1.48$. 
That makes the reverse shock less efficient to compress the beam matter (compared to the case of a shock 
with a lower polytropic index $\Gamma=4/3$), and
the region between the reverse shock and contact discontinuity then extends over
$\Delta X_{3}(t) = 0.0225  t$.

When we compare these two simulations, which differ in Lorentz factor of the beam, namely 100 versus 10,
these tests show the dual numerical difficulty when the Lorentz factor increases: (1) the compression rate enhances
and (2) the distance between the forward shock and contact discontinuity and between
the reverse shock and contact discontinuity decreases.

The last test is a variant of the first, with Lorentz factor of the beam 100 and transverse velocity 
$v_{4, y}=0.01$. Introducing a transverse velocity in the beam increases the
numerical difficulty to resolve the contact discontinuity with a jump in both density and transverse speed. 
The simulation is done with $256$ cells at the lowest grid level, and we 
allow for $12$ levels  on the spatial range $-0.1<x<10$, the result at $t=9.5$ is shown in Fig.~\ref{LabelFig_GRBTest100Tr}.
As a result of the interaction, the transverse velocity in the beam material shocked by the reverse shock, increases to 
$v_{3,y}=0.015$, which is a purely relativistic effect and depends both
on the relativistic kinematics as on the state of the matter \citep{Ponsetal00}. That
mechanism could be of interest in the refresh shocks for GRB afterglows, due to the effect of the blastwave meeting a sudden (wind) termination shock
in the CircumBurst Medium density profile \citep{Meliani&Keppens07}. In more than 1 dimension, such mechanism will contribute to the shocked
shell spreading with a speed higher than the sound speed.

In these tests, it became clear that in the case when both the forward and reverse shock
are relativistic, the constant polytropic index that should be used in simulations that do not have varying effective
polytropic index is in fact $4/3$. Even if
the external medium in these tests is cold, the important polytropic index is of the
shocked fluids. In other cases when the density ratio $n_{4}/n_{1}> \gamma_{4}$ is very
high, the reverse shock is near-Newtonian and the effective polytropic index increases to approach $5/3$ in the shocked medium between
reverse shock and contact discontinuity. 
Moreover, with a mildly-relativistic forward
or mildly relativistic reverse shock, as in the case with a beam Lorentz factor 
$\gamma_{4} \le 10$, the effective polytropic index in swept up ISM matter is higher than
$4/3$ while in the shocked beam matter the effective polytropic index can be mildly-relativistic 
as in Fig.~\ref{LabelFig_GRBTest10} where $\Gamma_{\rm eff}\sim 1.48$. For all cases, it is more convenient 
to use the new EOS.

\begin{acknowledgements}
We acknowledge financial support from the Netherlands Organization for Scientific 
Research, NWO-E grant 614.000.421, and from the FWO, grant G.027708, and computing resources supported by
NCF. Part of the computations made use of the VIC cluster at K.U.Leuven.
\end{acknowledgements}

\bibliographystyle{aa}

\end{document}